\documentclass[english,10pt]{article}
\usepackage{epsfig,amssymb,amsbsy}
\usepackage{amstext}
\usepackage{amsmath}
\usepackage{amssymb}
\usepackage{amsfonts}
\usepackage{color} 
\usepackage{siunitx}

\usepackage[round,numbers,sort&compress]{natbib}

\usepackage{biophysj}

\definecolor{Red}{rgb}{0.9,0.0,0.1}
\definecolor{Blue}{rgb}{0.1,0.0,0.9}

\begin{document}

\title{\bf
Feedback Mechanism for  Microtubule Length Regulation by 
 Stathmin Gradients}

\author{
\parbox{0.95\textwidth}{
\centering
Maria Zeitz, Jan Kierfeld\\[1ex]
{\footnotesize
Physics Department, TU Dortmund University,\\
44221 Dortmund, Germany\\
}
}
}

\date{\today}

\maketitle

\begin{abstract}
  We formulate and analyze a theoretical model for the regulation of 
  microtubule (MT) polymerization dynamics by
  the signaling proteins Rac1 and stathmin. In cells,  the MT
  growth rate is inhibited by cytosolic stathmin, 
 which, in turn, is inactivated by Rac1. 
  Growing MTs activate Rac1 at the cell edge, which closes
  a positive feedback loop. 
  We investigate both tubulin sequestering and catastrophe promotion as 
   mechanisms for MT growth inhibition by stathmin.
  For a homogeneous stathmin concentration in the absence of Rac1, we find 
   a switch-like regulation of the MT
  mean length by stathmin. For constitutively active Rac1 at the cell edge,
  stathmin is deactivated locally, which 
   establishes a spatial gradient of active stathmin. 
 In this gradient,  
  we find a stationary bimodal MT length distributions
 for both mechanisms of 
   MT growth inhibition by stathmin.
  One subpopulation of the bimodal length distribution
 can be identified with 
 fast growing and  long  pioneering MTs 
    in the region near the cell edge, which have been 
  observed experimentally. 
 The feedback loop is closed through Rac1 activation by MTs. 
  For   tubulin sequestering by stathmin,
  this  establishes a bistable  switch with two stable states:
  one stable state  corresponds to 
upregulated MT mean length and bimodal MT length distributions, 
i.e., pioneering MTs; the other stable state corresponds to 
an interrupted feedback with short MTs. 
Stochastic effects as well as external perturbations 
can trigger switching events. 
  For catastrophe promoting stathmin  
   we do not find bistability.
\end{abstract}


\section{Introduction}
 
Microtubules (MTs), an essential part of the cytoskeleton of
eukaryotic cells, are involved in  many cellular processes
such as 
 cell division \citep{Mitchison2001},
  intracellular positioning processes \citep{Dogterom2005} such 
as  positioning of 
 the cell nucleus \citep{Daga2006} or chromosomes during
mitosis, establishing of cell polarity \citep{Siegrist2007}, 
and  regulation of
cell  length \citep{Picone2010}. 
In all of these 
processes, the MT cytoskeleton has to be able to 
change shape and adjust the MT length distribution
by polymerization and depolymerization.
MT polymerization and 
depolymerization  also plays a crucial role in the
constant reorganization of the cytoskeleton of motile cells such as 
fibroblasts \citep{WatermanStorer1999}
or cells growing into polar shapes such as neurons 
\citep{Dehmelt2003}.
In motile cells, protrusion forces are often generated 
by the  actin lamellipodium at the cell edge, but MTs 
 interact with the actin cytoskeleton and actively 
participate in the regulation of motility \citep{WatermanStorer1999}. 
As a result, the MT cytoskeleton shape 
has  to adjust to changing 
cell shapes during locomotion.

The fast spatial reorganization of MTs is based on the dynamic
instability: phases of elongation by polymerization
 are stochastically interrupted by
catastrophes which initiate phases of fast depolymerization; fast
depolymerization terminates stochastically in a rescue event followed
again by a polymerization phase \citep{Mitchison1984}.

Regulation of MT length is crucial for the MT cytoskeleton to change 
shape. 
MT length regulation by depolymerases and polymerases 
such as kinesin-8 or XMAP215, which 
directly bind to  the MT, has been studied both 
experimentally (see Refs.\ \citep{Howard2007,Tolic-Norrelykke2010} 
for reviews) and theoretically 
\citep{Brun2009a,Tischer2010,Reese2011,Melbinger2012,Johann2012}.
Here, we want to explore and analyze 
 models for cellular MT length regulation by the 
signaling proteins Rac1 and stathmin, which
do not directly associate with MTs but 
are localized at the cell edge or in the cytosol, respectively.

Experiments have shown that dynamic MTs  participate
in regulation mechanisms at the lamellipodium of protruding cells
through interaction with Rac1
\citep{WatermanStorer1999,Kraynov2000,Wittmann2001}. 
Rac1 is a signaling molecule that controls actin dynamics and 
is essential for cell motility \citep{Kjoller1999}.   
 It is a GTPase of the Rho family which 
has been found to be active (phosphorylated) 
at the edge of protruding cells
\citep{WatermanStorer1999,Kraynov2000} as it becomes 
membrane-bound in its active state \citep{Moissoglu2006}. 
Rac1 activation at the cell edge has been shown 
 to be correlated with  MT polymerization 
 \citep{WatermanStorer1999}.
Therefore, it has been suggested that polymerizing MTs 
play an important role in activating Rac1  at the cell edge
\citep{Wittmann2001}.
The activation of Rac1 by MTs could involve their 
guanine-nucleotide-exchange factors. 
For the following, we will assume that 
Rac1 is activated by contact of MTs with the cell edge.

MTs are also
targets of cellular regulation mechanisms, which affect their dynamic
properties \citep{Athale2008}.  The dynamic instability of MTs enables
various regulation mechanisms of MT dynamics. 
In vivo, various MT-associated proteins 
have been found that either stabilize or destabilize MTs 
by direct interaction with the MT lattice, and
regulate MT dynamics both spatially and temporally \citep{Desai1997}.

 MT polymerization can also be regulated 
by the  soluble protein stathmin, which  diffuses freely in the
cytosol and inhibits MT polymerization \citep{Cassimeris2002}.
 The mechanism of MT inhibition by stathmin is 
still under debate \citep{Gupta2013} with the discussion focusing
on two mechanisms \citep{Cassimeris2002,Gupta2013}: 
\begin{enumerate}
\item 
   Stathmin inhibits MT growth via  sequestering of  free tubulin. 
One mole of active (nonphosphorylated)
 stathmin binds two moles of free tubulin and thereby lowers the local tubulin
 concentration
 \citep{Carlier2007,Jourdain1997,Curmi1997,Howell1999}. Consequently, 
    the growth velocity of the MT is suppressed and the
  catastrophe rate increased.
\item
 At high pH values, stathmin does not affect the 
growth velocity but only increases the MT catastrophe rate \citep{Howell1999}, 
possibly by 
direct interaction with the MT lattice \citep{Gupta2013}.
\end{enumerate}

Stathmin can be regulated by 
 deactivation upon phosphorylation \citep{Cassimeris2002}. 
One pathway of stathmin
regulation is the Rac1-Pak pathway, where Rac1 deactivates 
stathmin through the
intermediate protein Pak \citep{Wittmann2004,Watanabe2005}. 
Similar to Rac1, Pak has also been found to be localized in the 
leading edge of motile cells \citep{Sells2000}. 
Because the active form of Rac1
is situated at the cell edge, this introduces a spatial gradient of
stathmin phosphorylation and thereby stathmin tubulin interaction
\citep{Niethammer2004}.

\begin{figure}[h!]
\begin{center}
\includegraphics[width=0.98\linewidth]{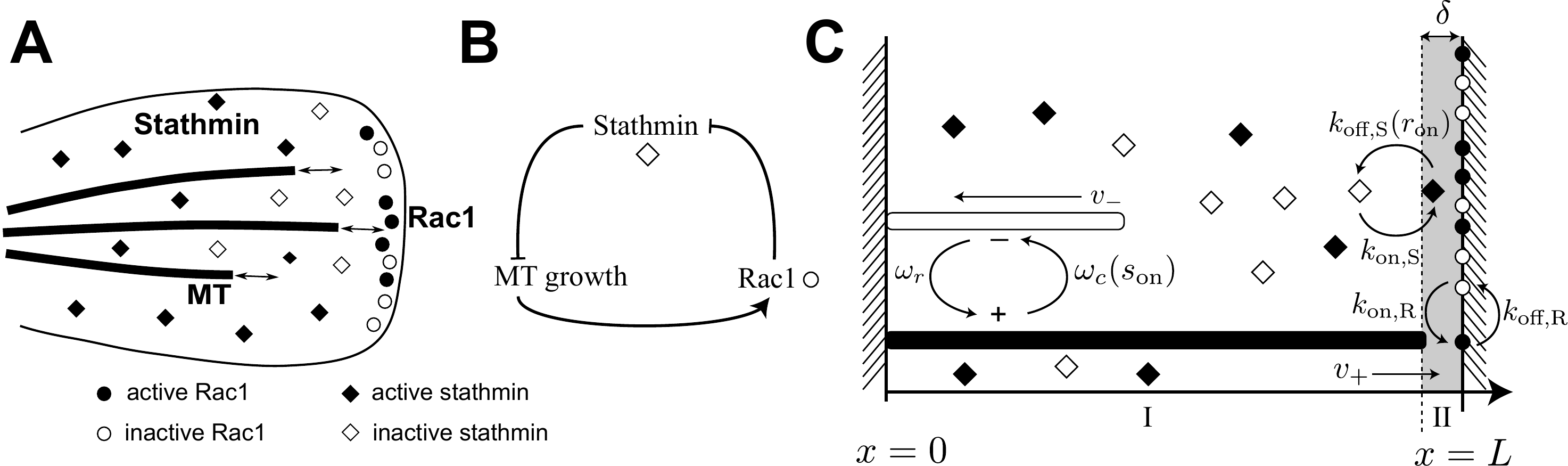}
\caption{ 
(A) Sketch of the feedback loop for MT growth regulation by Rac1 and stathmin
  in a cell. 
(B) Abstract scheme of the feedback loop.
(C)  Sketch of the one-dimensional model. The MT switches between growing
  ({\it solid bar}) or  shrinking state ({\it open bar}) with rescue 
  and catastrophe rates
  $\omega_{r}$ and $\omega_{c}$, respectively.  
  Rac1 proteins ({\it circles}) are located at the cell edge  $x=L$
  in active or inactive form ({\it solid} and {\it open circles},
  respectively).
 Rac1 is  activated
  with a rate $k_{\text{on,R}}$ only if the MT enters the grey shaded region II
  of size $\delta$ representing the cell edge, and
  deactivated at all times with the rate $k_\text{on,R}$. 
  Stathmin
  deactivation only takes place in the shaded region II with a rate
  proportional to the fraction of active Rac1 $k_\text{off,S}r_\text{on}$. 
  The
  local concentration of active stathmin inhibits  MT growth  
  by tubulin sequestering, which  reduces
   the MT growth velocity $v_+$ locally,  or by direct catastrophe promotion. 
  }
\label{fig:skizze}
\end{center}
\end{figure}

Thus, there is a positive feedback loop of MT
regulation \citep{Kuntziger2001,Wittmann2003,Li2005}, 
which consists of 
the activation of localized Rac1 by polymerizing 
MTs at the cell edge, the inhibition of 
cytosolic stathmin by active Rac1 via the Rac1-Pak pathway at the cell 
edge, and, finally, 
the inhibition of MT polymerization by stathmin, which sequesters 
tubulin or promotes catastrophes. The overall 
result is a  positive feedback loop consisting of one 
   positive (activating)  and  a doubly negative (inhibitory)
interaction (Fig.\ \ref{fig:skizze}, A and B).
Polymerizing MTs are an essential 
part of this feedback loop. 
Therefore, this feedback system
represents a  seemingly novel type of spatially organized 
 biochemical network, which could give 
rise to new types of spatial organization \citep{Dehmelt2010}.
In this article, we will formulate and analyze  a 
 theoretical model for this feedback loop  for 
tubulin-sequestering stathmin and for purely catastrophe-promoting stathmin.
Using this model we address the question how the feedback loop 
regulates the MT length.

The general structure of the described   positive feedback mechanism
 suggests two hypotheses for MT regulation, which we investigate within 
our model:
\begin{enumerate}
\item
The positive  feedback  increases MT growth and, thus, MT length.
\item
 Non-linearities in the positive feedback loop 
can give rise to a bistable switching 
between states of inhibited and increased MT growth.
\end{enumerate}

Regarding hypothesis 1, we will confirm that the regulation 
loop enhances MT growth. We will also show 
that polymerizing MTs  as one part of the loop result in a particular 
spatial organization of the feedback loop with
 {\it bimodal} MT length distributions.
This allows us to  successfully explain
the occurrence of pioneering MTs
as they have been observed in Waterman-Storer and Salmon
  \citep{WatermanStorer1997} and Wittmann et al.\ \citep{Wittmann2003}.
Such pioneering   MTs  grow into the leading edge of migrating 
cells and exhibit a decreased catastrophe 
frequency. In the experiments in  Wittmann et al.\ \citep{Wittmann2003},
 this pioneering behavior could be promoted by 
introducing constitutively active Rac1 and suppressed by introducing 
constitutively inactive Rac1 into the cells indicating that the 
mechanism underlying the pioneering MTs involves regulation by Rac1.
Our model  shows  there is a window of  
stathmin concentrations, for which
the MT length distribution is 
  bimodal and consists of two subpopulations:
 fast growing,  long  ``pioneering'' MTs near the cell edge;
  and  collapsed MTs with
   suppressed growth rates distant from the cell edge.
This bimodality is caused by a gradient in  stathmin phosphorylation.

Regarding hypothesis 2, we will show that 
our model exhibits bistability only if stathmin inhibits MT 
growth by tubulin sequestering.
The stathmin activation gradient, which  upregulates MT growth 
near the cell edge, represents
one stable state of the switch; complete stathmin deactivation with 
shorter MTs the other state. If stathmin acts by 
promoting  MT catastrophes, we do not find bistability.

\section{Methods}	

We formulate an effective model in which   we focus on MT
regulation through Rac1 and stathmin; we do not model 
the Rac1 activation pathway through MTs at the cell edge explicitly 
 (which requires most likely the action of additional 
guanine-nucleotide-exchange factors), 
and we do not model the pathway of 
stathmin deactivation by Rac1 explicitly (which also involves Pak and, 
most likely, other signaling molecules). 
The following experimental facts are included:
 MT growth is inhibited by active stathmin; we consider 
both tubulin sequestering and catastrophe promotion 
as  inhibition mechanisms. 
Dynamic MTs
activate localized Rac1 upon contacting the cell edge. 
Activated Rac1, in turn, inhibits stathmin at the cell edge.
Our effective model then reduces to  an amplifying positive feedback loop, 
which is composed of one positive and
 two inhibitory interactions (see Fig.\ \ref{fig:skizze} B).

We describe this regulation mechanism
 as a modified reaction-diffusion process that includes 
 directed MT polymerization and
approach the problem both by particle-based 
stochastic simulations and  analytical mean-field calculations. 
For simplicity, we consider 
one spatial dimension, i.e., regulation in a narrow box
of length $L$ along 
the $x$-axis of the growing MT. The model  includes $m$ independent MTs
 confined within the box, Rac1 proteins, which are localized at the
 right boundary of the box representing the cell edge, 
and stathmin proteins, which diffuse freely along the box 
(Fig.\ \ref{fig:skizze} C).
Our model for MT regulation 
is characterized by a number of parameters, for many of  which  experimental
information is already available, which we collected in 
Table \ref{tab:literature} \citep{Drechsel1992, Gildersleeve1992, Walker1988, Janson2004,Pryer1992,Nakao2004,Haeusler2006,Haeusler2003,Curmi1997,Redeker2000,Honnappa2003, Amayed2002}.

\begin{table}
\caption{Literature values for parameters}
\begin{tabular}{lcr}
\hline
Parameter & Values & References\\
\hline \hline
Microtubule & &\\
\hline
Growth velocity $v_{+}$  &7-80 nm/s&
    \citep{Drechsel1992, Gildersleeve1992, Walker1988, Janson2004} \\
Shrinking velocity $v_{-}$ &0.18-0.5 $\mu$m/s &
    \citep{Drechsel1992, Gildersleeve1992, Walker1988} \\
Effective dimer length  $d$ & $8/13 {\rm nm} \simeq 0.6 {\rm nm}$\\
Tubulin association rate $\omega_\text{on}=\kappa_\text{on}[T_0]$  &62-178 1/s &
    \citep{Walker1988} \\
 ~~for $[T_0]= 7-20 {\rm \mu M}$ &&\\
Rescue rate $\omega_{r}$ & 0.05-0.5 1/s& 
   \citep{Walker1988, Pryer1992,Nakao2004}\\
Catastrophe rate $\omega_{c}=\frac{1}{a+bv_{+}}$ & 
      $b=\SI{1.38e10}{s^{2}m^{-1}}$& 
   \citep{Janson2003b}\\
  & $a=\SI{20}{s}$& \\
\hline
Rac & & \\
\hline 
Intrinsic hydrolysis rate $k_\text{off,R}$& 0.0018-0.0023/s& 
     \citep{Haeusler2006,Haeusler2003}\\
Hydrolysis rate $k_\text{off,R}$ & 0.039/s&\citep{Haeusler2006}\\
 ~~in the presence of GAP&  & \\
\hline
Stathmin  & & \\ \hline
Diffusion coefficient $D$& 13(COPY)-73(RB3) $\mu \text m^{2}$/s& 
    \citep{Carlier2007,Niethammer2004}\\
Stokes radius $R_{s}$ & 33-39 \AA& 
     \citep{Schubart1987, Curmi1997,Redeker2000}\\
Activation gradient length scale $\chi_{s}$&  4-8 $\mu$m &
      \citep{Niethammer2004}\\
Phosphatase activity $k_\text{on,S}$ & 0.3-0.7 1/s&  
      \citep{Niethammer2004}\\
Sequestering eq.\  constant $K_{0}=1/K_{D}$&1.43-169/$\mu\text{M}^{2}$& 
    \citep{Carlier2007,Niethammer2004,Honnappa2003,Curmi1997, Amayed2002}\\
Catastrophe promotion constant $k_c$ & 
  0.002 ${\rm s^{-1}\mu M^{-1}}$ & \citep{Howell1999}\\
\hline
\end{tabular}
\label{tab:literature}
\end{table}

\subsection{Single microtubule in a box}

The MT dynamics in the presence of the dynamic instability is described in
terms of a stochastic two-state model \citep{Verde1992,Dogterom1993}: 
In the growing state, a MT
polymerizes with an average velocity $v_+$. The MT stochastically switches
from the growing state  to a  shrinking state  with the catastrophe rate
$\omega_c$. In the shrinking state, it rapidly depolymerizes with an average
velocity $v_-$. With the rescue rate $\omega_r$, the MT stochastically 
switches back  to the growing state.

The stochastic time evolution of an ensemble of 
independent MTs, growing along
the $x$-axis, can be described by two coupled master equations for the
probabilities $p_+(x,t)$ and $p_-(x,t)$ of finding a MT with length $x$ at
time $t$ in a growing or shrinking state \citep{Dogterom1993}: 
\begin{align}
 \partial _t p_+(x,t)&=-\omega_c p_+(x,t)+\omega_r p_-(x,t)
            -v_+\partial _x p_+(x,t),	\label{p+}\\
  \partial _t p_-(x,t)&=\omega_c p_+(x,t)-\omega_r p_-(x,t)
            +v_-\partial _x p_-(x,t).	\label{p-}
\end{align}
We confine the MT within the one-dimensional cell 
of length $L$ by  reflecting boundary conditions:
shrinking back to zero length gives rise to  a forced rescue and 
contacting the boundary at $x=L$ gives rise to  an instantaneous
 catastrophe event. This corresponds to 
\begin{equation*}
v_+p_+(0,t)-v_-p_-(0,t)=v_+p_+(L,t)-v_-p_-(L,t)=0.
\end{equation*}
Forced rescue at $x=0$ is equivalent to immediate renucleation 
 of MTs, which are anchored at a MT organizing center.
Our model does not apply to unanchored treadmilling MTs.

For a $x$-independent  catastrophe rate $\omega_c$,
the steady state 
 probability density function of finding a MT of length $x$
 can be determined analytically \citep{Tischer2010,Zelinski2012},
as
\begin{equation}
	p(x)=p_{+}(x)+p_{-}(x)= 
    N \left(1+\frac{v_+}{v_-}\right)e^{x/\lambda},
 \label{eq:pvonx}
\end{equation}
with a normalization
\begin{equation*}
N^{-1}=\lambda\left(1+{v_{+}}/{v_{-}}\right)\left(e^{L/\lambda}-1\right),
\end{equation*}
where $\lambda$  is a characteristic length parameter, 
\begin{equation}
	\lambda\equiv\frac{v_+ v_-}{v_+\omega_r-v_-\omega_c}.		
   \label{lambda}
\end{equation}
If $\lambda<0$, the average length loss after catastrophe 
exceeds the average length gain;
 if $\lambda>0$, the average length gain exceeds 
the average length loss. 
The resulting average MT length  is given by
\begin{equation}
	\langle x_\text{MT} \rangle = \int_{0}^{L}x p(x)dx
   =\frac{L}{1-e^{\lambda/L}}-\lambda. 
\label{eq:xMT}
\end{equation}

In the growing state, GTP-tubulin dimers are attached to any of the 13
protofilaments with the rate $\omega_{\text{on}}$, which is proportional
to the local concentration of free 
 GTP-tubulin $[T]$, 
$\omega_{\text{on}} = \kappa_{\text{on}}[T]$. 
GTP-tubulin dimers are detached with the rate
$\omega_{\text{off}}$.
 The resulting growth velocity is
obtained by multiplication with the effective tubulin dimer size $d
\approx \SI{8}{nm}/13\approx\SI{0.6}{nm}$,
\begin{equation}
 v_+ =
    \left(\kappa_{\text{on}}[T]-\omega_{\text{off}}\right) d
\label{v+}.
\end{equation}

In the classical model of the MT catastrophe mechanism,
 the catastrophe rate $\omega_{c}$ is
 determined by the hydrolysis dynamics of GTP-tubulin with 
each loss of the stabilizing GTP-cap due to hydrolysis 
 causing a catastrophe.  Experimental results 
\citep{Janson2003b} show that the average time 
 spent in the growing state, $\langle \tau _{+}\rangle=1/\omega_c$,
 is a linear function of the growth velocity $v_{+}$,
\begin{equation}
	\omega_{c}=\frac{1}{a+bv_{+}},	
\label{omegac}
\end{equation}
with constant coefficients $a=\SI{20}{s}$  and 
$b=\SI{1.38e10}{s^{2}m^{-1}}$.
All of our results will be 
 robust against the choice of the catastrophe model
(see the Supporting Material).

\subsection{Rac model}

Rac1 (in the following called ``Rac'') proteins are small GTPases,
whose  activity is regulated by GTP-binding (activation) and
intrinsic hydrolysis (deactivation).  
Rac proteins are localized
 at the edge of protruding cells
\citep{Moissoglu2006,WatermanStorer1999,Kraynov2000}, and 
polymerizing MTs   activate Rac  at the cell edge
\citep{Wittmann2001}.

We model Rac proteins as pointlike objects situated
 at the boundary $x=L$, 
which  can exist in two states, activated or deactivated.  
We  denote the fraction of activated  Rac 
by $r_\text{on}$.
A boundary region $x\in[L-\delta, L]$
of small size $\delta$  represents the cell edge
(region II in Fig.\ \ref{fig:skizze} C); because the details of 
the Rac activation are not known, we include $\delta\ll L$
 as a reaction distance:
Rac activation  takes place with a 
constant rate $k_\text{on,R}$  and only if one of the $m$ 
  MT contacts the right boundary region in its growing state.
The deactivation of Rac, on the other hand, happens 
independently of the
MT growth state  with the  constant
rate $k_{\text{off,R}}$.

The chemical kinetics of Rac at the cell edge can be  described by 
\begin{equation}
  \frac{\partial r_\text{on}}{\partial t}=
     p_\text{MT}k_\text{on,R}(1-r_\text{on}) - k_\text{off,R}r_\text{on},
 \label{eq:diffrac}
\end{equation}
where the Rac activation rate is analogous to 
a second order reaction with the mean MT cell edge contact probability
(for contact in the growing state) 
\begin{equation}
   p_\text{MT}(t) = m \int_{L-\delta}^L p_+(x,t),
\label{eq:pMT}
\end{equation}
playing the role of 
a MT concentration, and  the Rac deactivation rate is first-order. 
Equations \eqref{eq:diffrac} and \eqref{eq:pMT} 
describe the Rac kinetics at the 
mean-field level  neglecting temporal correlations between 
 MT cell edge  contacts  and Rac number fluctuations.
In a stationary state,
Eq.\ \eqref{eq:diffrac} gives a  Rac activation level
\begin{equation}
   r_\text{on}=   \frac{1}{1+ k_\text{off,R}/p_\text{MT}k_\text{on,R}}.
 \label{eq:ronstat}
\end{equation}
Because Rac activation requires MT contact, changes in the 
activation rate $k_\text{on,R}$ (for which there is no 
experimental data yet)  can always be compensated by changing
the number of MTs $m$, as it is apparent from Eq.\ \eqref{eq:ronstat}.

\subsection{Stathmin model}

Stathmin is a soluble protein that  diffuses freely in the
cytosol and inhibits MT growth. 
We consider two possible inhibition 
mechanisms \citep{Cassimeris2002}: 1) Tubulin sequestering,
and 2) catastrophe promotion. 
MT growth inhibition is turned off by
stathmin phosphorylation 
through the Rac-Pak pathway \citep{Wittmann2004,Watanabe2005}.

\subsubsection{Gradient in  stathmin activation}

The interplay of  stathmin  diffusion and
phosphorylation  by 
active Rac at the cell edge  
establishes a spatial gradient of stathmin
activation.

Stathmin proteins are modeled as pointlike objects 
in  an active or inactive state. 
Unlike Rac proteins, stathmin proteins  diffuse
freely within the whole simulation box $x\in [0;L]$ assuming concentration
profiles 
$S_\text{on}(x)$ in the active  dephosphorylated state,
which can inhibit MT growth,
and $S_\text{off}(x)$ in the inactive phosphorylated state
(with a total stathmin concentration $S_\text{tot}(x) 
  = S_\text{on}(x)+S_\text{off}(x)$). 
We assume the diffusion coefficient $D$ to be constant and equal
for both states.  Switching between
 active  and  inactive state
 is a stochastic process with
 rates $k_{\text{on,S}}$ and $k_{\text{off,S}}$, respecively.

Rac acts through Pak as a stathmin kinase.
We describe  stathmin  deactivation 
by Rac at the cell edge $x=L$ as  simple second order
reaction with a rate proportional to $r_\text{on}$
(note that including Michaelis-Menten like enzyme kinetics for stathmin
deactivation does not alter the
results qualitatively).

We assume that the phosphatase responsible for 
stathmin activation is homogeneously
 distributed within the cytosol
 such that stathmin dephosphorylates
 everywhere within the box with the constant rate
 $k_\text{on,S}$ \citep{Brown1999}.  
 The  distribution of deactivated
 stathmin in the one-dimensional box is then 
 described by a reaction-diffusion equation 
(see the Supporting Material for details).

In the steady state, we find a 
 stathmin  activation gradient with a profile
\begin{equation}
 \frac{S_\text{on}(x)}{S_\text{tot}}=
      1- 2A \cosh(x/\chi_S),
     \label{eq:son}
 \end{equation}
 which decreases with increasing $x$ towards the cell edge $x=L$
 because stathmin is deactivated at the cell edge by Rac. 
The  
 characteristic decay length for the stathmin activation gradients is 
 given by 
\begin{equation*}
\chi_S=\sqrt{{D}/{k_\text{on,S}}}
\end{equation*}
 and arises from the competition of stathmin reactivation in the bulk 
 of the box and diffusion.
The  integration constant  
\begin{equation}
    A= \frac{1}{2} \frac{r_\text{on} k_{\text{off,S}}}{ (D/\delta\chi_S)
         \sinh(L/\chi_S) + 
      (r_\text{on}k_{\text{off,S}}+ k_{\text{on,S}}) \cosh(L/\chi_S) }.
 \label{eq:A}
 \end{equation} 
depends on the 
 degree of stathmin deactivation by Rac at the cell edge $x=L$
 and, thus, on the fraction $r_\text{on}$ 
 of activated Rac (see the Supporting Material).

Experimentally, a characteristic stathmin 
gradient  length scale $\chi_{s} =   4-8 {\rm \mu m}$ 
has been measured   using 
fluorescence resonance energy transfer 
 with fluorescent stathmin fusion proteins (COPY) \citep{Niethammer2004}. 
Values for the diffusion constant $D=13-18  {\rm \mu m^2}/s$
of the  fluorescent COPY proteins \citep{Niethammer2004}  are consistent 
with activation or dephosphorylation
rates $k_\text{on,S} =  0.3-0.7 {\rm 1/s}$ by phosphatase activity.

\subsubsection{Tubulin-sequestering stathmin}

One possible pathway for stathmin to inhibit MT growth is 
sequestering of free 
tubulin,  which  lowers the MT growth velocity 
$v_{+}$ by
decreasing the local concentration of free tubulin. It also alters the
catastrophe rate $\omega_{c}$ via the growth velocity, as described by Eq.\
\eqref{omegac} \citep{Curmi1997}.  

One 
active stathmin protein sequesters two tubulin proteins
\citep{Carlier2007,Jourdain1997,Curmi1997,Howell1999},
\begin{equation}
 2 T + 1 S\rightleftharpoons ST_2,
\label{eq:TSreaction}
\end{equation}
  and  the growth velocity $v_{+}$ depends on the concentration 
 $[T]$ of free tubulin as described in Eq.\ \eqref{v+}.
Solving the chemical equilibrium equations
we find the normalized concentration of free tubulin $t\equiv[T]/[T_0]$ 
(normalized by the total tubulin concentration $[T_0]=[T]+2[ST_2]$)
as a function of normalized active stathmin 
 $s_\text{on} \equiv S_\text{on}/[T_0]$,
\begin{align}
t(s_\text{on})&= \frac{1}{3}\left[ 1-2   s_\text{on}
  +\frac{-3+k  (1-2 s_\text{on})^2}{k \alpha(s_\text{on})}
         + \alpha(s_\text{on})\right] ~~\mbox{with}
\nonumber\\
\alpha(s_\text{on})&\equiv 
  \left[ (1-2 s_\text{on})^3+ (9/ k)(1+s_\text{on})+ \right.
   \nonumber\\
 &~~~\left.   +3  \sqrt{(3/k ^3) \left(1+k ^2 (1-2 s_\text{on})^3+k  
     \left(2 +10  s_\text{on}-s_\text{on}^2\right)\right)}   \right]^{1/3},
\label{tn(x)}
\end{align}
where $k\equiv K_0[T_0]^2$ denotes the normalized equilibrium constant
of the reaction Eq.\ \eqref{eq:TSreaction}.
The resulting curve $t(s_\text{on})$ is strongly non-linear and 
agrees well with experimental results on the amount $1-t$ of 
bound tubulin in Jourdain et al.\ \citep{Jourdain1997}.

\begin{figure}[h!]
\begin{center}
\includegraphics[width=0.6\linewidth]{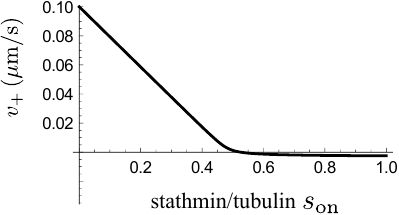}
\caption{
  Growth velocity $v_{+}$ as a function of the normalized concentration 
  of  active stathmin $s_\text{on} = S_{\text{on}}/{[{T_0}]}$ as given 
 by inserting ${[{T}]}= {[{T_0}]} t(s_\text{on})$
   according to Eq.\ \eqref{tn(x)}  into 
 Eq.\ \eqref{v+} for $k=K_0{[{T_0}]}^2=9400$ corresponding to
  $K_0 = 25 {\rm \mu M}^{2}$ and ${[{T_0}]} = 19.4 {\rm \mu M}$, 
  see Table \ref{tab:parameter}. 
 The sharp crossover at 
$s_\text{on}\simeq 0.5$ is  a result of the tubulin sequestering
in a 1:2 complex.
}
\label{fig:stathminvelocity}
\end{center}
\end{figure}

Inserting  the result $[T]= [T_0] t(s_\text{on})$ 
in the linear equation \eqref{v+} for the 
growth velocity, 
we also obtain a strongly 
nonlinear dependence of the growth velocity on active stathmin
(see Fig.\ \ref{fig:stathminvelocity}). 
Assuming that tubulin-stathmin association is fast 
compared to the other processes, a
local concentration  $s_{\text{on}}(x)$
of active stathmin also 
 gives rise to a 
local concentration $[T](x)=[T_0] t(s_\text{on}(x))$ 
of free tubulin, and Eq.\  \eqref{v+} determines
the local growth velocity: 
\begin{equation}
  v_+ = v_+([T](x)) = v_+([T_0] t(s_\text{on}(x))).
\label{eq:v+profile}
\end{equation}

\subsubsection{Catastrophe-promoting stathmin}

Another  possible pathway for stathmin to inhibit MT growth is by directly 
promoting catastrophes, possibly via
direct interaction with the MT lattice \citep{Gupta2013}. 
This pathway might be more relevant at high pH, whereas tubulin sequestering 
dominates at lower pH values \citep{Howell1999}. For high pH,  the data of 
Howell et al.\ \citep{Howell1999} are consistent with a linear increase
of the catastrophe rate with the concentration of active stathmin,
\begin{equation}
	\omega_{c}(S_\text{on}) = \omega_{c}(0) +  
                  k_c S_\text{on} = \omega_{c}(0) +  
                  k_c [T_0]s_\text{on},
\label{eq:omegacSon}
\end{equation}
with a catastrophe promotion constant $k_c = 0.002 {\rm s^{-1}\mu M^{-1}}$.
An activation gradient of stathmin $S_\text{on}(x)$ then 
 gives rise to a  local MT catastrophe rate $\omega_c(S_\text{on}(x))$. 

We will use both alternatives, 
 the   catastrophe promotion pathway of stathmin
and  the sequestering pathway in our model, and compare the resulting 
MT growth behavior.

\subsection{Stochastic simulations}

The model outlined in the previous section is the basis 
for our one-dimensional stochastic simulations
(employing equal time steps $\Delta t=0.001 {\rm s}$).
 We model the MTs as straight  polymers with continuous length $x$, and 
Rac and Stathmin distributions via discrete particles,
which can be activated and deactivated according to the 
rates specified above. The $N_S$ stathmin particles can  
diffuse freely within the 
simulation box, whereas the $N_R$ 
Rac particles are localized at the cell edge region of size $\delta$. 
Details of the simulation are described in the Supporting Material.

In Table \ref{tab:parameter} 
we present our choice of parameters to perform
simulations. We assume a  tubulin concentration 
$[T_0]= 19.4 {\rm \mu M}$
We choose a simulation box of length $L=10 {\rm \mu m}$ with a 
cell edge region of size $\delta = 20 {\rm nm}$, where Rac can 
be activated by MTs. 
In the Supporting Material it is shown 
that our results are qualitatively similar for larger  
lengths $L$.

\begin{table}
\caption{Parameter values used for simulation}
\begin{tabular}{lcc}
\hline
Description & Parameter & Value/Reference\\
\hline\hline
Time step & $\Delta t$ & 0.001 s\\
System length & $L$ & 10 $\mu$m\\
Cell edge region & $\delta$ & 0.02 $\mu$m\\
\hline
Microtubule & & \\
\hline 
Tubulin concentration &  $[T_0]$ &  $19.4 {\rm \mu M}$ \\
Effective dimer length & $d$ & $8/13 {\rm nm} \simeq 0.6 {\rm nm}$\\
Growth velocity ($S_{\text{on}}=0$) &$v_{+}$  &0.1 $\mu$m/s \\
Shrinking velocity &$v_{-}$ & 0.3 $\mu$m/s\\
Tubulin association rate & $\omega_\text{on}=\kappa_\text{on}[T_0]$  & 173/s
        \citep{Walker1988} \\
Dissociation velocity & $v_\text{off}$ & $3.6 {\rm nm/s}$ 
           \citep{Janson2004}  \\
Rescue rate& $\omega_{r}$ &0.1/s\\
Catastrophe rate ($S_{\text{on}}=0$) & $\omega_c$ & 0.0007/s \\
Number of tubulin dimers & $N_{T}$&10000\\ 
  ~~corresponding to a volume & $V$ &  $0.86 {\rm \mu m}^3$\\
\hline
Rac & & \\
\hline 
Number of Rac molecules & $N_{R}$ & 1000\\
Activation rate & $k_\text{on,R}$& 5/s\\
Deactivation rate & $k_\text{off,R}$& 0.002/s\\
\hline
Stathmin & & \\
\hline 
Activation rate & $k_\text{on,S}$& 1/s\\
Deactivation rate & $k_\text{off,S}$& 300/s\\
Diffusion coefficient & $D$&  15 $\mu\text m^{2}$/s\\
Sequestering eq.\  constant  & $K_0$& 25/$\mu\text{M}^{2}$\\
         & $k = K_0[T_0]^2$ &  9400  with $[T_0]=19.4 {\rm \mu M}$\\ 
Catastrophe promotion constant & $k_c$ &
  0.002 ${\rm s^{-1}\mu M^{-1}}$\\
\hline
\end{tabular}
\label{tab:parameter}
\end{table}

\section{Results and Discussion}

\subsection{Interrupted feedback for 
 constitutively active stathmin in the 
 absence of Rac} 
\label{OpenF}

In a first step, we  investigate how 
constitutively active stathmin alters the MT
dynamics, which is equivalent to interrupting the feedback loop by 
removing Rac ($r_\text{on}=0$) such that 
 stathmin cannot be deactivated  and is homogeneously
distributed, 
$S_\text{tot} = S_\text{on} = {\rm const}$, due to diffusion. 
Because Rac is absent, all results are 
 independent of the membrane contact probability 
 of MTs, and we can limit the MT number to $m=1$.

\begin{figure}[h!]
\begin{center}
\includegraphics[width=0.95\linewidth]{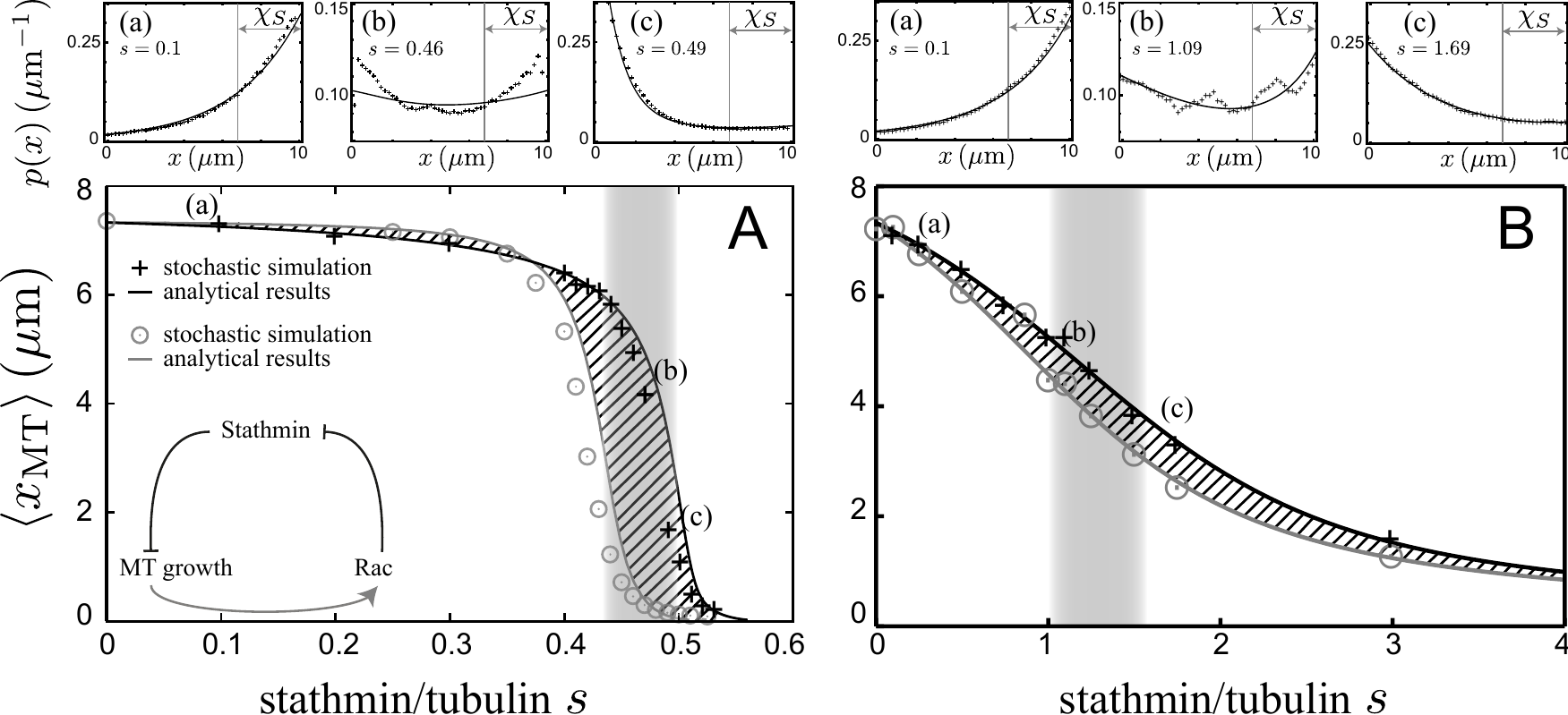}
\caption{ 
 Stochastic simulation data ({\it data points}) 
  and analytical master equation  results  ({\it solid lines})
   for the
  mean MT length $\langle x_{\text{MT}}\rangle$ 
   as a function  of stathmin/tubulin
    $s=S_{\text{tot}}/{[{T_0}]}$.
  We compare the system with 
  constitutively active Rac ($r_\text{on}=1$, {\it solid lines} and 
 {\it crosses}) 
  to the  system without Rac ($r_\text{on}=0$, {\it shaded lines} and 
   {\it circles}) both for tubulin-sequestering 
   stathmin (A) and catastrophe-promoting stathmin (B).
   ({\it Hatched area}) Possible MT length gain by Rac regulation.
   ({\it Shaded area}) Region, 
   in which MTs exhibit bimodal length distributions
  for constitutively active Rac.
  ({\it Insets a-c}) Corresponding 
  MT length distributions for three particular 
  values of $s$ with (a) $s<s_\lambda$, (b) $s=s_\lambda$, and (c)
  $s>s_\lambda$.
}
\label{fig:MittelwerteallesRac}
\end{center}
\end{figure}

For  homogeneously distributed stathmin,  the 
MT growth rate $v_+$ and the catastrophe rate $\omega_c$ are  homogeneous 
both for tubulin-sequestering and for catastrophe-promoting stathmin.
Then, we can calculate the  MT length distribution  
from  the analytical solution \eqref{eq:pvonx}  for a position-independent 
parameter $\lambda$  (see Eq.\ \eqref{lambda}).
 We  find good agreement 
between the analytical result Eq.\ \eqref{eq:xMT} for the 
MT mean length 
and stochastic simulation results  as a function of 
the stathmin/tubulin
$s \equiv  S_\text{tot}/[T_0]$ 
($s=s_\text{on}$ for constitutively active stathmin) 
both for tubulin-sequestering and catastrophe-promoting
stathmin (Fig.\ \ref{fig:MittelwerteallesRac}). 
Deviations are due
 to stochastic fluctuations of the local stathmin concentration because of 
   diffusion.

Our main finding  is that stathmin
regulates MT growth in a switchlike manner
both if stathmin sequesters tubulin and if stathmin promotes catastrophes. 
For tubulin-sequestering stathmin the switch is much steeper,
and we can define two
stathmin concentrations,  $S_\lambda$ and $S_v$, 
which characterize the steepness of
the switch.  At the critical concentration $S_\lambda$ 
the characteristic length
parameter $\lambda$  (see Eq.\  \eqref{lambda}), changes sign, i.e., 
 $\lambda^{-1}(S_\text{tot}=S_{\lambda})=0$. 
The condition $\lambda^{-1}=0$ results in a flat 
MT length distribution \eqref{eq:pvonx} and, thus, an alternative 
definition of $S_\lambda$ is 
$\langle x_\text{MT} \rangle(S_\text{tot}=S_{\lambda}) = L/2$.

Above the critical stathmin concentration  $S_{\lambda}$, 
we find the MT length
distributions to be negative exponentials.
At the second, higher critical  concentration $S_v$ 
the MT growth velocity  $v_{+}$ (see Eqs.\ \eqref{v+} and \eqref{tn(x)})
changes sign, i.e., $v_{+}(S_\text{tot}=S_{v})=0$.
Above $S_v$,
 the MT cannot grow at all and 
$\langle x_{\text{MT}} \rangle=0$ for $S\ge S_v$.
For catastrophe-promoting stathmin, the switch is much broader
because the MT growth velocity is not affected, formally resulting 
in an infinite $S_v$. A critical concentration $S_\lambda$ can be defined
in the same way as for tubulin-sequestering stathmin.

For tubulin-sequestering stathmin the 
 critical stathmin/tubulin
 $s_\lambda$ and $s_v$ are generally close to $0.5$  as a result 
of the 1:2 tubulin sequestering.
For catastrophe-promoting stathmin, 
 $s_\lambda$ approaches the 
 $[T_0]$-independent limit
$s_\lambda \approx \kappa_{\text{on}} d \omega_r/v_-k_c = 0.91$,
which is  $\gg 0.5$ (for more details, see the 
Supporting Material).
The values $s_\lambda$ and $s_v$ are only weakly $[T_0]$-dependent,
 which motivates using  the  stathmin/tubulin
 $s=s_\text{on}$ as a control parameter.

We find that constitutively active stathmin
regulates MT growth in a switchlike manner for a fixed tubulin 
concentration  $[T_0]$ as characterized by the critical concentration
$S_\lambda$, which is approximately linear in 
the tubulin concentration $[T_0]$ for both models of stathmin action. 
Vice  versa, we  conclude that, at  a fixed stathmin 
concentration $S_\text{tot}$, the tubulin concentration $[T_0]$
can 
regulate MT growth in a switchlike manner with a critical  concentration
$[T_0] \sim s_\lambda/S_\text{tot}$,
which can be controlled by the stathmin concentration $S_\text{tot}$.
Without stathmin regulation,  a critical tubulin concentration 
$[T] = \kappa_\text{on}/\omega_{\text{off}}$ for MT growth at the 
plus end exists, but
this concentration is  fixed by the rate constants
of the growth velocity $v_+$ appearing in Eq.\ \eqref{v+}.
Moreover, the switchlike behavior becomes sharpened by stathmin regulation.

In vivo, 
the picture will be complicated if  
additional  populations of treadmilling MTs exist
apart from the end-anchored MTs, which we consider here.
For tubulin-sequestering stathmin, 
treadmilling MTs will act as additional tubulin buffer.
Upon sequestering by stathmin 
the length of treadmilling MTs will adjust  such that the 
concentration of free tubulin is maintained at the critical 
 concentration for treadmilling.

For a system which contains constitutively active stathmin and no Rac, 
the growth velocity $v_{+}$ is homogeneous,
and we always expect exponential MT length distributions as in Eq.\ 
\eqref{eq:pvonx}.
In particular, we do not find  bimodal MT length 
distributions in the absence of Rac
(see also Supporting Material).

\subsection{Interrupted feedback for constitutively active Rac} 

Now we reinclude Rac into our analysis of the feedback mechanism.
We start with  constitutively active Rac ($r_\text{on}=1$), which 
is analogous to interrupting the 
feedback between MTs and Rac by making  the level of Rac activation 
 independent of the  cell edge  contacts of the MTs.
Thus, we can limit the MT number to $m=1$ also in this section.

Constitutively active Rac establishes a gradient of active stathmin, which is
constant in time, which can be quantified by 
the analytical result from Eq.\ \eqref{eq:son} 
 with $r_\text{on}=1$.
 As shown in the Supporting Material, it
agrees very well with stochastic simulation results. 
The  decreasing 
profile $S_\text{on}(x)$ of activated stathmin is essential for 
MT length regulation because it leads to an increasing 
 growth velocity $v_x(x)$ (for tubulin-sequestering stathmin) and/or a
decreasing catastrophe rate $\omega_c(x)$ (for tubulin-sequestering and
catastrophe-promoting stathmin),  which modulates the 
MT length distribution.

We can calculate the MT length distribution starting from 
the negative
stathmin  gradient 
\begin{equation*}
s_\text{on}(x) =  S_\text{on}(x)/[T_0] = 
     s (S_\text{on}(x)/S_\text{tot})
\end{equation*}
  (with the total 
   stathmin/tubulin $s = S_\text{tot}/[T_0]$) as given by 
Eq.\ \eqref{eq:son}.
For tubulin-sequestering stathmin, this results in an
increasing tubulin concentration 
\begin{equation*}
 [T](x) = [T_0] t(s_\text{on}(x))
\end{equation*}
through Eq.\ \eqref{tn(x)}, which gives rise to both an
increasing  growth velocity $v_+(x)$ according to Eq.\
(\ref{eq:v+profile}) and a decreasing  catastrophe rate 
$\omega_c=\omega_c(v_+(x))$.
For  catastrophe-promoting stathmin, only the 
catastrophe rate $\omega_c=\omega_{c}(S_\text{on})$ becomes decreasing
via eq.\ \eqref{eq:omegacSon}. 
For both stathmin models, this results in an  
  increasing inverse  characteristic length parameter 
\begin{equation*}
  \lambda^{-1}(x) = 
  \omega_r/v_- - \omega_c(x)/v_+(x)
\end{equation*}
(see Eq.\ \eqref{lambda}). 
The master equations
(Eqs.\ \eqref{p+} and \eqref{p-}) then give the  steady-state 
probability distribution for the MT length \citep{Zelinski2012}, 
\begin{align}
p(x)&=p_+(x)+p_-(x)=
   N\left(1+\frac{v_-}{v_+(x)}\right)
    \exp\left(\int_0^x dx'/\lambda(x')\right),
\label{eq:pgrad}
\end{align}
with a normalization 
\begin{equation*}
N^{-1}=\int_0^Ldx\left(1+\frac{v_-}{v_+(x)}\right)
  \exp\left(\int_0^x dx'/\lambda(x')\right).
\end{equation*}

In contrast to the system without Rac,  position-dependent MT growth 
velocity and/or catastrophe rates  can give rise to more complex
MT length distributions 
because 
the  decreasing profile $S_\text{on}(x)$ of active stathmin
gives rise to an  
 inverse parameter $\lambda^{-1}(x)$, which  is an increasing function 
of $x$ for both stathmin models. 
The most interesting situation then 
arises if $\lambda^{-1}(x)$ changes 
its sign on $0<x<L$. Then, we find  $\lambda(x)<0$ or MTs shrinking on
average for small $x$ and $\lambda(x)>0$ or MTs growing on average 
for large $x$. 
The  decreasing profile $S_\text{on}(x)$ of active stathmin  thus leads 
to  unstable  MT growth because 
it promotes growth of long MTs and shrinkage 
  of short MTs. 
In fact, we find bimodal MT length distributions for this case
 as a result of the negative gradient or decreasing  profile  of active
stathmin according to Eq.\ \eqref{eq:son}  (and see 
Fig.\ \ref{fig:MittelwerteallesRac} {\it insets (b)}).

These bimodal MT length distributions correspond to two 
populations of MTs:  long fast-growing  MTs and 
short collapsed  MTs.
The subpopulation of long fast-growing MTs resembles the experimentally
observed  long  pioneering microtubules 
\citep{WatermanStorer1997,Wittmann2003}.
Pioneering   MTs  have been observed to 
grow into the leading edge of migrating 
cells and exhibit a decreased catastrophe 
frequency. Because of the reduced growth velocity by 
 deactivated stathmin, these are exactly the properties of the 
subpopulation of long MTs.
We conclude that the bimodal MT length distribution 
occurring at high active Rac fractions can explain 
the phenomenon of pioneering MTs.

It is instructive to compare the impact of 
the decreasing  active stathmin profile 
$S_\text{on}(x)$  on the MT length distribution with the 
 effect of  an external force $F(x)$, which opposes MT growth 
and increases with $x$, for example, as for an elastic obstacle 
 \citep{Zelinski2012,Zelinski2013}.
Whereas an opposing force $F(x)$, which increases with $x$, 
gives rise to a decreasing growth velocity  $v_+(x)$ and a  decreasing
parameter $\lambda^{-1}(x)$, it is found that a decreasing 
profile  $S_\text{on}(x)$  always gives rise to an
increasing parameter $\lambda^{-1}(x)$.
This difference leads to very different behavior.
For an opposing force $F(x)$, the MT length $x_\text{MT}$ with 
$\lambda^{-1}(x_\text{MT})=0$ represents a  stable equilibrium:
shorter MTs grow on average, whereas longer MTs shrink. Therefore, 
we find a pronounced maximum in the MT length distribution 
around $x\sim x_\text{MT}$ \citep{Zelinski2012,Zelinski2013}.
For stathmin-regulated MT growth, a decreasing profile $S_\text{on}(x)$ 
leads to an  unstable equilibrium as outlined above. 
As a result, we find a  bimodal MT length distribution with a 
 minimum  in the MT length distribution  
around the MT length $x_\text{MT}$ with $\lambda^{-1}(x_\text{MT})=0$.

In Fig.\  \ref{fig:MittelwerteallesRac}, we show the average length 
$\langle x_\text{MT}\rangle$ of the MT
 as a function of the stathmin/tubulin 
$s=S_\text{tot}/[T_0]$ for the two subsystems 
with constitutively active Rac ({\it black lines} and {\it symbols}) 
and without Rac ({\it shaded lines} and {\it symbols}) 
for both tubulin-sequestering 
and catastrophe-promoting  stathmin.
The  black lines corresponds to the  analytical steady state 
solution following the calculation outlined before;
 black  data points 
are results from   fully  stochastic simulations. 
Also for active Rac, the total  stathmin
concentration regulates MT growth in a switch-like manner, and
we can define a critical 
stathmin concentration $S_\lambda$ from the condition 
$\langle x_\text{MT} \rangle(S_\text{tot}=S_{\lambda}) = L/2$.
The corresponding values 
are slightly higher than  in the absence of Rac.
We find bimodal MT length distributions for stathmin 
concentrations close 
to the critical value $s_\lambda$ because 
the condition  $\langle x_\text{MT} \rangle = L/2$ implies that 
there exists a
 MT length $x_\text{MT}$ with $\lambda^{-1}(x_\text{MT})=0$.

The deviation between the stochastic simulation results 
and the analytical master equation
solution  in Fig.\  \ref{fig:MittelwerteallesRac}
is due to fluctuations in the local concentration of active
stathmin.
Also in  stochastic simulations, 
we find bimodal MT  length distributions and, thus, 
pioneering MTs  for stathmin/tubulin $S_\text{tot}/[T_0]$ in the 
 gray-shaded area  in Fig.\ \ref{fig:MittelwerteallesRac}
around the critical value $s_\lambda$.

\subsection{Closed Feedback}	

Now we consider the full system with closed feedback,
where  the activation of 
Rac proteins  depends on
cell edge  contacts of the MTs and, therefore, also 
changes with the number of MTs $m$ in the system.

For the closed feedback loop, we expect 
 that the average MT length will lie in
between our results without Rac ($r_\text{on}=0$)
 and for full Rac activation ($r_\text{on}=1$), 
as indicated by the
hatched areas  in Fig.\ \ref{fig:MittelwerteallesRac}.
For a closed regulation feedback loop, the  system can vary the 
MT length  between these two bounds, for example, by increasing 
the number $m$ of MTs and, thus, the Rac activation level. 
The MT length is most sensitive to 
regulation in the vicinity of the switch-like behavior, i.e., for 
stathmin/tubulin 
$S_\text{tot}/[T_0]\sim s_\lambda$.

\begin{figure}[h!]
\begin{center}
    \includegraphics[width=0.99\linewidth]{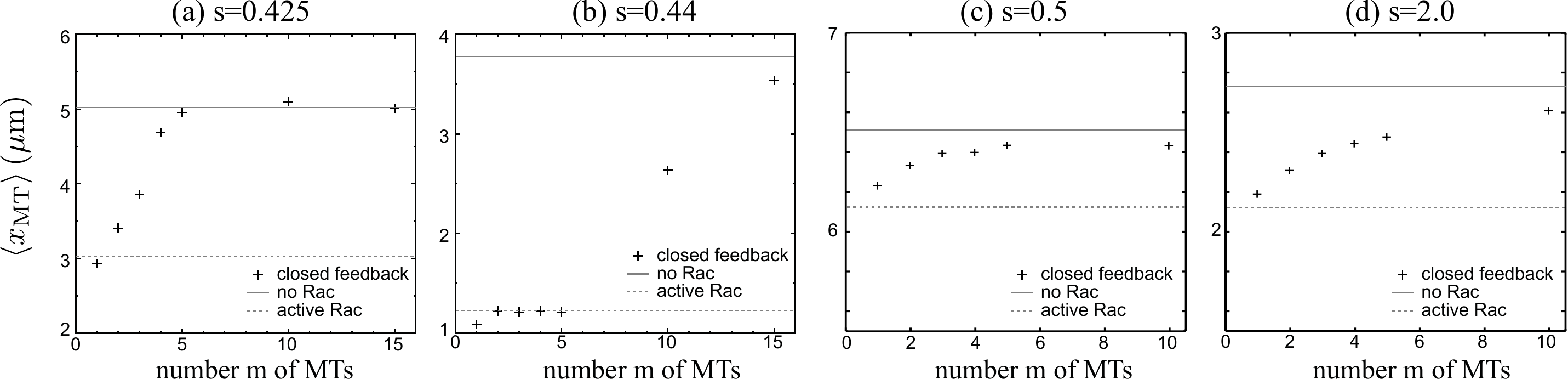}
\caption{
   The average MT length 
     $\langle x_\text{MT}\rangle$ as a function of the number of MTs $m$ 
    in a system with closed feedback for 
    tubulin sequestering stathmin (a and b) and catastrophe 
  promoting stathmin (c and d) at 
   different stathmin/tubulin 
    $s = S_{\text{tot}}/{[{T_0}]}$ values: (a) $s=0.425$, 
   (b) $s=0.44$, (c) $s=0.5$, (d) $s=2.0$.  
  ({\it Horizontal solid} and {\it dashed  lines}) Mean MT length 
  for systems without Rac ($r_{\text{on}}=0$) 
   and constitutively active Rac ($r_{\text{on}}=1$), 
  respectively. 
    }
	\label{fig:closedFB}
\end{center}
\end{figure}

Using fully stochastic simulations 
we investigate the dependence of the mean MT length 
  $\langle x_\text{MT}\rangle$
on the number $m$ of MTs  for
stathmin/tubulin  $S_\text{tot}/[T_0]\sim s_\lambda$
 ($s \approx 0.4-0.5$ for 
tubulin-sequestering stathmin and
$s\approx 1.0-1.5$ for catastrophe-promoting stathmin).
The results are shown in  Fig.\ \ref{fig:closedFB}.
For a closed feedback loop, the
mean MT length increases with 
the number $m$ of MTs between the two bounds
given by  a system without Rac and a system with complete Rac activation,
 because  the strength of the feedback increases with $m$
via the cell edge contact probability  as given by Eq.\ \eqref{eq:pMT}. 
If the MT  contact probability is sufficiently high, the system will
stabilize long MTs by  Rac activation. 
For tubulin-sequestering stathmin, the average MT
length increases linearly only above a  critical number $m_{c}$ of MTs. 
This critical number 
$m_{c}$ increases with the stathmin concentration. 
For catastrophe-promoting stathmin, there is no critical 
number of MTs necessary, and the length increases 
gradually as a function of MT number $m$.

We can rationalize these results by the bifurcation behavior 
of the stationary state of the feedback system at the mean-field level,
which is qualitatively different for tubulin-sequestering stathmin
and catastrophe-promoting stathmin.
For a fixed mean 
concentration $r_\text{on}$ of activated Rac, we can follow the 
same steps as in the previous section to calculate the
resulting stationary stathmin profile \eqref{eq:son} with 
an arbitrary $0\le r_\text{on}\le 1$ in Eq.\ \eqref{eq:A} for 
the parameter $A$ and, then, the  
steady state
probability distribution $p(x)$ for the MT length  (Eq.\ \eqref{eq:pgrad})
given the velocity profile in Eq.\ \eqref{eq:v+profile}. 
From $p(x)$ we obtain the MT cell edge 
contact probability $p_\text{MT}$ in the stationary state 
using Eq.\ \eqref{eq:pMT}.
All in all, we can calculate  $p_\text{MT}$  
 from  any given fixed Rac activation level $r_\text{on}$.
On the other hand, for a closed feedback loop, the MT 
contact probability $p_\text{MT}$ feeds back and 
determines the  Rac activation level 
$r_\text{on}$ via Eq.\ \eqref{eq:ronstat} in the stationary state. 
At the stationary state of the closed feedback loop both relations 
have to be fulfilled simultaneously, which is only possible at certain 
{\em fixed points} of the feedback system.

\begin{figure}[h!]
\begin{center}
\includegraphics[width=0.95\linewidth]{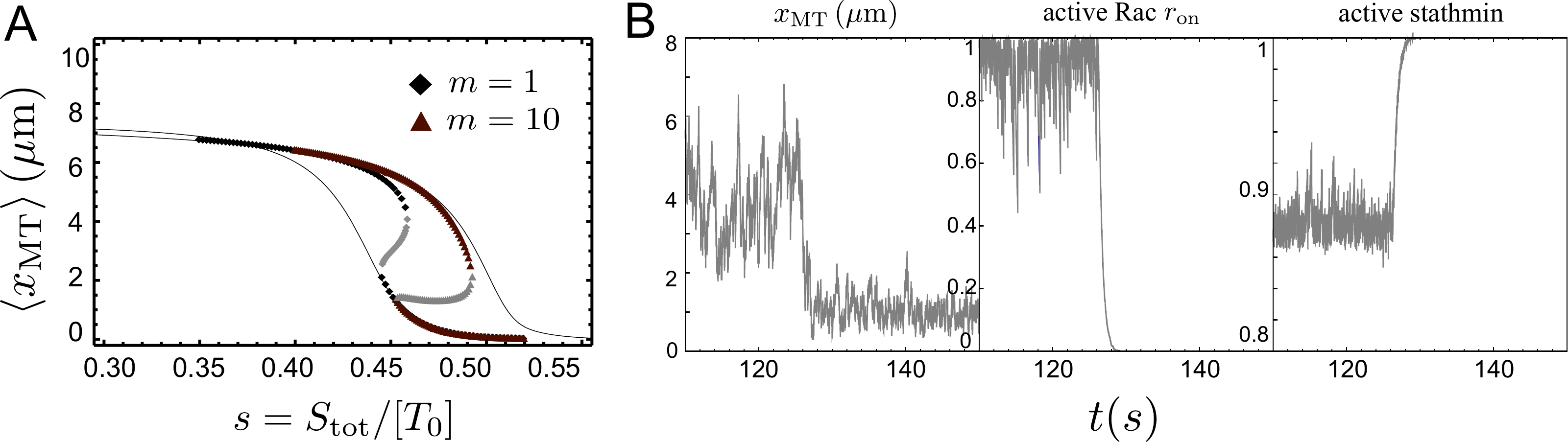} 
\caption{
   (A)
  Average MT length $\langle x_\text{MT} \rangle$ 
   at  the fixed points as a function of 
  stathmin/tubulin     $s=S_{\text{tot}}/{[{T_0}]}$
   for tubulin-sequestering stathmin ({\it diamonds}, $m=1$; {\it triangles},
   $m=10$;  {\it gray}, unstable fixed point). 
  ({\it Lines})  $\langle x_\text{MT} \rangle$ 
  for a system without Rac  ({\it lower line}, $r_\text{on}=0$)
   and for constitutively active Rac ({\it upper line}, $r_\text{on}=1$). 
  (B) Time evolution of a feedback system with 
    $m=1$ MT for tubulin-sequestering stathmin and
  $s=0.44$.
   Plots show MT length, the fraction $r_\text{on}$ of active Rac 
  and the total fraction of active stathmin 
  $L^{-1}\int_0^L dx  S_{\text{on}}(x)/S_{\text{tot}}$ 
   as a function of time. 
  Around $t \sim 130 {\rm s}$ the system switches from the 
   high active Rac fixed point to an interrupted feedback fixed point. 
}
	\label{fig:closedFBdynamics}
\end{center}
\end{figure}

For tubulin-sequestering stathmin,
 there are up to three fixed points, which 
exhibit two saddle-node bifurcations
typical for a bistable switch
(see the Supporting Material for more details).
For small stathmin concentrations
there is 
only a single fixed point at a high level  $r_\text{on}\approx 1$ 
of active Rac corresponding 
to upregulated MT growth resulting in a 
large mean MT  length  $\langle x_{\text{MT}}\rangle$ and bistable 
length distributions, i.e., pioneering MTs.
At this  fixed point, $\langle x_{\text{MT}}\rangle$  is described by
the black curve in Fig.\ \ref{fig:MittelwerteallesRac}.
For high stathmin concentrations 
there exists only a single fixed point at low active 
Rac $r_\text{on}\approx 0$ corresponding 
to an interrupted  feedback loop  with short MTs.
At this  fixed point, $\langle x_{\text{MT}}\rangle$  is given by
the shaded curve  in Fig.\ \ref{fig:MittelwerteallesRac}.
The upper critical  stathmin/tubulin  increases with 
 the MT number $m$. 
In between 
we find a  hysteresis with three fixed points,
one of which is unstable, corresponding to a bistable switch behavior.
The hysteresis loop widens for increasing MT numbers $m$,
which is also why the  critical  $m_c$ rises  as a function of $s$ (see 
Fig.\ \ref{fig:closedFB}).
This gives also rise to hysteresis in the mean MT length, as shown in 
Fig.\ \ref{fig:closedFBdynamics} A.

We conclude that for tubulin-sequestering stathmin 
closure of the positive  feedback mechanism leads to a 
 bistable switch, which  locks either  into  a high concentration
of active Rac with   upregulated MT growth 
and pioneering MTs or into a low concentration of active Rac
with   quasi-interrupted feedback and short MTs. 
This bistable behavior requires a certain threshold
 value $m k_\text{on,R}> 1.9/{\rm s}$ such  
that the feedback is strong enough. As of this writing, 
there is no  experimental data for $k_\text{on,R}$;  
values  $k_\text{on,R}< 1.9/{\rm s}$ would require more than one MT 
to trigger bistability.
In our model this bistable switching behavior occurs even in the absence 
of a nonlinear Michaelis-Menten kinetics for stathmin phosphorylation
because the dependence of the contact probability $p_\text{MT}$
on the Rac activation level 
$r_\text{on}$  is strongly nonlinear for tubulin-sequestering stathmin.
The observed bifurcation is a {\em stochastic bifurcation} 
(of P-type)
\citep{Zakharova2010} in the sense that the MT length distribution 
undergoes a qualitative change at the bifurcation.

This  bistability is also observed in our  stochastic simulations
for tubulin sequestering stathmin (see Fig.\ \ref{fig:closedFBdynamics} B), 
where stochastic fluctuations 
give rise to  switching from the high active 
Rac fixed point ($r_\text{on}\approx 1$) 
to an interrupted feedback at 
 low active Rac ($r_\text{on} \approx 0$).
If we start in a state at the high Rac fixed point and 
the time  between successive cell-edge contact becomes comparable 
to the time constants $k_\text{off,R}$ for Rac deactivation and 
$k_\text{on,S}$ for stathmin reactivation, the feedback can be 
interrupted by stochastic fluctuations. Then the system
 falls into the low Rac fixed point,  as can be seen in 
 Fig.\ \ref{fig:closedFBdynamics} around  $t\sim 130s$. 
Because the MT length distribution is exponentially decreasing 
at the low Rac fixed point, the reverse event of 
switching back to a high Rac value is improbable. 
Stochastic  switching events as shown 
in Fig.\ \ref{fig:closedFBdynamics} for $m=1$ MT become increasingly 
rare for large numbers $m$ of MTs, which 
effectively stabilizes the fixed point of  upregulated MT growth.

The main source of stochasticity is the growth dynamics of
MTs and the resulting time-dependence of the contact 
probability: for small numbers $m$,
the constant average value $p_\text{MT}$ 
is only established as average over 
many contacts with relatively long times between successive contacts.
Therefore, we expect the mean-field results to become correct only 
for large MT numbers $m$. 
For small $m$, we find pronounced stochastic effects.
Furthermore,  the gradual linear increase 
of $\langle x_{\text{MT}} \rangle$  above the critical MT number $m_c$ 
in Fig.\ \ref{fig:closedFB} is the result 
of averaging bistable switching over many realizations.

For catastrophe-promoting stathmin, there is only 
a single fixed point, which explains  the gradual 
upregulation of MT growth 
by the MT number $m$. 
The reason for the absence of any bistability is the 
strictly linear dependence of the  inverse  characteristic length parameter 
\begin{equation*}
  \lambda^{-1}(x) = 
  \omega_r/v_--\omega_c(x)/v_+(x)
\end{equation*}
on the concentration of active stathmin  (see Eqs.\ 
\eqref{eq:omegacSon} and \eqref{lambda}). 
This gives rise to a quasi-linear dependence of the contact
 probability $p_\text{MT}$ on the Rac activation level 
$r_\text{on}$ for catastrophe-promoting stathmin.

Our theoretical findings 
show the following
\begin{enumerate}
\item
 Rac activation in the cell-edge region 
 can establish the feedback, 
 increase MT growth, and trigger MT pioneering behavior
in accordance with experimental findings in Ref.\ \citep{Wittmann2003}
and, 
\item
 Surprisingly,  simply increasing the number $m$ of 
 participating  MTs
can also upregulate  growth of these MTs, inasmuch as
this increases the feedback strength and  
the level of active Rac.
\end{enumerate}


\section{Conclusion}\label{Discussion}

We formulated and investigated a theoretical model for
the influence of a doubly negative feedback mechanism
on MT growth involving 
the signaling proteins Rac1 and stathmin. 
MTs activate Rac1 in the cell edge region; activated 
localized Rac1 inhibits stathmin, which freely diffuses in the cytosol
and is an inhibitor for MT growth.
We studied and compared two models for the MT inhibiting 
effect of stathmin, tubulin sequestering and catastrophe promotion. 
The resulting 
 positive feedback loop is of particular interest because of 
the prominent role spatial organization: MTs grow along a particular
direction and stathmin organization along this direction is essential.

For high concentrations of active Rac1, our model 
produces  a stathmin activation 
gradient, which leads to an increased mean MT length and
  bimodal MT length distributions. This  explains the phenomenon 
of pioneering MTs that have been observed experimentally 
\citep{WatermanStorer1997,Wittmann2003}. 
We find bimodal  MT length distributions both for tubulin-sequestering
  and for catastrophe-promoting stathmin.
Our model shows that 
the stathmin activation gradient requires a  local activation of 
Rac1 at the cell edge. Local Rac1 activation can happen by 
MT contacts for closed feedback,
but an  external local Rac1 activation should  also  trigger 
a stathmin activation gradient and, thus, pioneering MTs 
in accordance with experimental observations \citep{Wittmann2003}. 
  Localized activation of other kinases 
which inactivate 
stathmin could have a similar effect, whereas stathmin-targeting
kinases, which are active throughout the cytoplasm, will not give 
rise to stathmin activation gradients or  pioneering MTs.

Both for constitutively active and inactive Rac1 we find a 
switchlike dependence of the mean MT length on the 
overall concentration ratio of stathmin to tubulin. 
We find such a  switchlike dependence 
both for tubulin-sequestering and catastrophe-promoting stathmin
with a characteristically steeper switch for tubulin-sequestering stathmin.
This qualitative difference between tubulin-sequestering 
and catastrophe-promoting stathmin should be amenable to experimental 
testing, which might help to settle the yet-unresolved question 
of the MT inhibiting mechanism of stathmin.

The positive feedback mechanism involving Rac1 and stathmin 
allows to upregulate MT growth 
within the bounds set by  constitutively active and inactive Rac1. 
Based on our theoretical model, we predict a qualitatively different 
 effect of the positive feedback  for 
tubulin-sequestering and catastrophe-promoting stathmin.
For tubulin-sequestering stathmin,
we identify a bistable switch  with 
 two stable fixed points within a mean-field theory. 
One fixed point corresponds to high active Rac1, 
upregulated MT mean length and bimodal MT length distributions, 
i.e., pioneering MTs; the other fixed point corresponds to 
an interrupted feedback at low active Rac1 concentrations,
with short MTs. 
Stochastic fluctuations can give rise to spontaneous stochastic 
switching events, which we also observed in simulations.

Interestingly, we find bistable switching behavior even in the absence 
of a nonlinear Michaelis-Menten kinetics for stathmin phosphorylation.
The  bistability is due to nonlinear dependencies of the 
MT cell-edge contact probability on the active Rac1 concentration via 
the  free tubulin concentration and the stathmin concentration. 
For catastrophe-promoting stathmin, this nonlinearity is 
absent and we find a gradual increase of MT length by the 
positive feedback mechanism.
These qualitative differences between tubulin-sequestering 
and catastrophe-promoting stathmin in the theoretical model 
could give important 
indications for experiments in order to resolve the issue of the 
mechanism of MT growth inhibition by stathmin: experimental 
observations of bistability would clearly hint towards 
the tubulin sequestering mechanism.

We checked that all of our results are robust against
changes of the catastrophe model and the 
cell length $L$ (see the Supporting Material).
Based on our model, we suggest three mechanisms to influence 
the overall  MT growth behavior or the Rac1-induced upregulation of 
MT growth
in vivo or by external perturbation in experiments:
\begin{enumerate}
\item
  The overall MT length can be controlled via the total 
tubulin or stathmin concentrations. Rising  the tubulin concentration
or lowering the stathmin concentration upregulates
MT growth. 
\item 
Rac1 activation in the cell-edge region increases  MT growth and can 
trigger pioneering MTs. 
\item 
 Increasing the number $m$ of 
 participating  MTs
can also increase MT growth and  trigger pioneering MTs.
\end{enumerate}  
All of these mechanisms should be accessible in perturbation 
experiments. 
Moreover, our results on MT length distributions can be checked against
future quantitative length measurements on pioneering MTs.

Our results have  implications for the polarization of
the MT cytoskeleton as well.  
 In a population of MTs growing in different 
directions, the investigated positive feedback mechanism via 
Rac1 and stathmin regulation can select and amplify 
a certain MT growth direction, such as one triggered by a
 locally  increased
Rac1 concentration in the cell-edge region. 
This issue remains to be investigated in the future.

\section{Acknowledgments}

We thank Bj\"orn Zelinski, Susann El-Kassar and Leif Dehmelt 
for fruitful discussions. 
We acknowledge support by  the Deutsche Forschungsgemeinschaft 
(KI 662/4-1).

\section{Supporting Citations}

References \citep{Flyvbjerg1996,Li2009} appear in the 
Supporting Material.


\newpage
\setcounter{section}{0}
\setcounter{equation}{0}
\setcounter{figure}{0}
\renewcommand{\theequation}{{S\arabic{equation}}}
\renewcommand{\thefigure}{{S\arabic{figure}}}

\section*{Supporting Material}

 The Supporting Material contains
details of the simulation methods for the 
feedback system of polymerizing microtubules, 
localized Rac1 and cytosolic stathmin.
We also present additional model equations and 
 simulation results on the 
stathmin activation gradient.
We  show additional results on the interrupted feedback 
subsystem without 
Rac and constitutively active stathmin. 
For the full system with closed feedback,  we present  details of the 
theoretical bifurcation analysis.
Finally, we present additional results on the robustness
of our results with respect to  changes in the catastrophe model 
and the system length.

\section{Simulation methods}

The model described in the main text  is the basis 
for our one-dimensional stochastic simulation. 
We employ  a stochastic 
simulation with equal time steps $\Delta t=0.001 {\rm s}$, i.e., 
 the simulation time is discretized 
into equidistant discrete times  $t_{n}=n\Delta t$.  
 We model the MTs as straight  polymers with continuous length, and 
Rac and Stathmin distributions via discrete particles,
which can be activated and deactivated according to the 
rates specified in the main text. 
Stathmin particles can  diffuse freely within the 
simulation box, whereas Rac particles are localized at the cell-edge 
region.

In Table \ref{tab:parameter} in the main text 
we present our choice of parameters to perform
simulations. We assume a  tubulin concentration 
$[T_0]= 19.4 {\rm \mu M}$, and  we assign a volume 
$0.86 {\mu m}^3$ to our simulation box, which is equivalent to 
a total number of $N_T=10000$ tubulin dimers in the simulation box.
We choose a simulation box of length $L=10 {\rm \mu m}$ with a 
cell-edge region of size $\delta = 20 {\rm nm}$, where Rac can 
be activated by MTs.

\subsection{Microtubule simulation}

In order to simulate one single MT in a box, we use 
the following approximations:
\begin{enumerate}
\item[1)] 
The MT is rigid and grows and shrinks in only one direction, along the
$x$-axis. 
The MT tip position is described by a continuous variable 
$x_\text{MT}$. 
\item[2)]
The one-dimensional box has fixed rigid walls at $x=0,L$. 
At $x_\text{MT}=L$ (the cell edge), a growing MT
experiences an immediate  catastrophe event. For simplicity, 
we assume that at $x_\text{MT}=0$, a shrinking MT starts to grow again.  
\item[3)]
Tubulin is not simulated explicitly. The interaction with stathmin is
implemented indirectly by a change of the growth velocity via the
concentration 
$[T]$ of free tubulin following Eq.\  \eqref{v+} in the main text.
The total number $N_T =10000$ of tubulin dimers 
 is only needed to calculate values of
 stathmin/tubulin  $s = S/[T_0] = N_S/N_T$ 
conveniently.
 \end{enumerate}

The MT is completely described by the position $x_\text{MT}$ of the
tip and its growth  state (growing or shrinking). 
 The switching of the MT state is described as a stochastic
event with the time constant probabilities for catastrophe
$p_{c}=\omega_{c}\Delta t$ and rescue $p_{r}=\omega_{r}\Delta t$ 
in each time step.
Whereas the rescue rate $\omega_r$ is always independent of the MT 
tip position $x_\text{MT}$, the catastrophe rate
 $\omega_{c}(x_\text{MT})$ becomes a function of $x_\text{MT}$
in the presence of a stathmin concentration profile as further explained
below.
 If the MT has grown in the last time step and 
 a random number $c\in [0;1]$ is smaller
than $p_{c}$, then the MT shrinks in the current time step, and if $c>p_{c}$
the MT continues to grow (and vice versa).  The new coordinate is
\begin{equation}
x_{\text{MT,}n+1}=x_{\text{MT,}n}+v_{+,n}\Delta t 
   \quad \text{or}\quad 
 x_{\text{MT,}n+1}=x_{\text{MT,}n}-v_{-}\Delta t
\end{equation}
for growing or shrinking respectively.  
At the boundary of the simulation box
 for positions $x_\text{MT}>L$ a catastrophe is enforced, $p_{c}=1$,
and for $x_\text{MT}<0$, we enforce immediate rescue,  $p_{r}=1$.  
The growth velocity
$v_{+,n}$ depends on the local  concentration of 
free tubulin at $x_{\text{MT,}n}$ 
via Eq.\ \eqref{v+} in the main text and, for tubulin-sequestering stathmin, 
on the local concentration 
of  active stathmin  via Eq.\ \eqref{tn(x)} in the main text.

In general, we perform simulations with $m$ MTs, in a sufficiently big ensemble
(in average 100 independent systems compose an ensemble). In order to minimize
fluctuations, the observables are measured in the steady state and 
averaged over  sufficiently large time intervals.


In order to measure the length distribution, i.e., the probability to 
 find a MT of length $x$ during the simulation, 
we divide the
box into bins of the length $\Delta L$ and typically use 
 $\Delta L=L/50$.

\subsection{Rac simulation}

The $N_{R}$ Rac proteins are described as point-like objects which are all
situated in the right boundary at $x=L$, which represents 
the cell edge. In each simulation 
time step $\Delta t$, each  Rac protein is 
deactivated with the constant probability 
$p_\text{off,R}=k_\text{off,R}\Delta t$.
If a Rac protein is active, we draw  a random number $c\in [0;1]$;
if $c$  is smaller
than $p_\text{off,R}$  it is deactivated in the following time step. 

In order to activate Rac, a
membrane contact of the MT is necessary. If one of the $m$ 
MTs is at the cell edge region, each Rac protein is activated 
with a probability $p_\text{on,R}=k_\text{on,R}\Delta t$ in each time step
$\Delta t$.
If one MT is at the position $x_\text{MT}\in [L-\delta; L]$ and
a random number  $c\in [0;1]$ is smaller than 
  $p_\text{on,R}$ the protein is activated in the next time step. 
 Since the
details of the Rac activation are not known, we include $\delta\ll L$ as a
reaction distance.

\subsection{Stathmin simulation}

The $N_S$ 
stathmin proteins are point-like objects which diffuse freely within the 
one-di\-men\-sional box $x\in[0,L]$.
The probability density function
for the distance $\Delta x$ that 
 a diffusing particle moves in a simulation time step $\Delta t$ 
is a Gauss distribution
\begin{equation}
p(\Delta x,\Delta t)= \frac{1}{\sqrt{4\pi D \Delta t}}
    e^{-\frac{\Delta x^{2}}{4D\Delta t}}.
\end{equation}
The mean square distance a particle moves during the time step
$\Delta t$ is $\langle \Delta x^2 \rangle = 2 D \Delta t$.
We determine 
 the new position of a stathmin protein (index $j$) as
\begin{equation}
x_{j,n+1}=x_{j,n}+d\sqrt{2 D \Delta t},
\end{equation}
where $d$ denotes a random number drawn from a standard normal distribution.
using a Box-Muller algorithm.  
The position $x_{j,n}\in[0;L]$ is limited to
the box  by reflecting boundary conditions.

We activate each
stathmin protein regardless of its  position within
the cell with the probability $p_\text{on,S}=k_\text{on,S}\Delta t$ in each
simulation time step $\Delta t$. 
This is implemented by comparison of $p_\text{on,S}$ with 
a random number $c\in [0;1]$.
The deactivation of a stathmin protein, 
on the other hand, only takes place if it is
positioned in the interval $[L-\delta;L]$. 
Then, it is deactivated with the probability
$p_\text{off,S}=r_\text{on}k_\text{off,S}\Delta t$, with $r_\text{on}$ being
the fraction of active Rac proteins at this time step.
Activation and deactivation are
 implemented by comparison of $p_\text{on,S}$ and $p_\text{off,S}$,
 respectively, with 
a random number $c\in [0;1]$.

We compare two mechanisms for the inhibition of MT growth by 
stathmin: (a) tubulin sequestering by stathmin resulting 
in reduced free tubulin concentrations, reduced MT growth 
velocities, and increased MT catastrophe rates and (b) a purely
MT catastrophe-promoting activity of stathmin.

For both mechanisms,  we
measure the local concentration 
$S_\text{on}(x)$ of active stathmin during the simulation,
in order to simulate the local impact of stathmin on the growth velocity 
of a MT with tip position $x$. 
In order to define and measure the local concentration 
$S_\text{on}(x)$ of active stathmin during the simulation, 
we divide the
box into bins of the length $\Delta L$ and count the active 
and inactive stathmin
proteins. Because we choose $\Delta L$ much smaller than the 
 stathmin gradient scale $\chi_S$, 
see \eqref{eq:appchi}  below, 
the total number of stathmin proteins is nearly constant in
each bin.  Typically, we choose $\Delta L=L/50$.

\subsubsection{Tubulin-sequestering stathmin}

For the tubulin-sequestering model of stathmin, 
we do not simulate the binding of tubulin dimers to stathmin
explicitly but assume that the binding to stathmin is fast 
compared to the other processes. Therefore, we can determine
the local concentration  $t(s_\text{on}(x))$ of free tubulin 
from the local concentration of active stathmin 
$s_\text{on}(x) =S_\text{on}(x)/[T_0]$ 
via the chemical equilibrium relation \eqref{tn(x)} for 
a fixed total
tubulin concentration $[T_0]$ corresponding a number $N_T=[T_0]V$ 
of tubulin dimers in a volume $V$. 
Once we know the  local concentration $[T](x)=[T_0] t(s_\text{on}(x))$ 
of free tubulin, Eq.\  \eqref{v+} in the main text determines
the local growth velocity, 
\begin{equation}
  v_+ = v_+([T](x)) = v_+([T_0] t(s_\text{on}(x)))
\end{equation}
(see Eq.\ \eqref{eq:v+profile} in the  main text).
This quasi-equilibrium relation  allows us to determine the local MT growth 
velocity $v_+=v_+(x)$ uniquely from the  local concentration 
$S_\text{on}(x)$ of active stathmin. 

Because the catastrophe rate $\omega_c$ is determined by $v_+=v_+(x)$ 
via relation (\ref{omegac}) in the main text, a local growth velocity 
also gives rise to a local catastrophe rate $\omega_c=\omega_c(v_+(x))$ 
in the tubulin-sequestering model of stathmin.

\subsubsection{Catastrophe-promoting stathmin}

For the catastrophe-promoting model of stathmin, 
the  local concentration 
$S_\text{on}(x)$ of active stathmin does not affect the MT
growth velocity $v_+$ but directly the catastrophe rate via the 
relation  \eqref{eq:omegacSon} in the main text. 
This gives rise to a local catastrophe rate 
$\omega_{c}= \omega_{c}(S_\text{on}(x))$.

\begin{figure}[h]
 \begin{center}
 \includegraphics[width=0.7\linewidth]{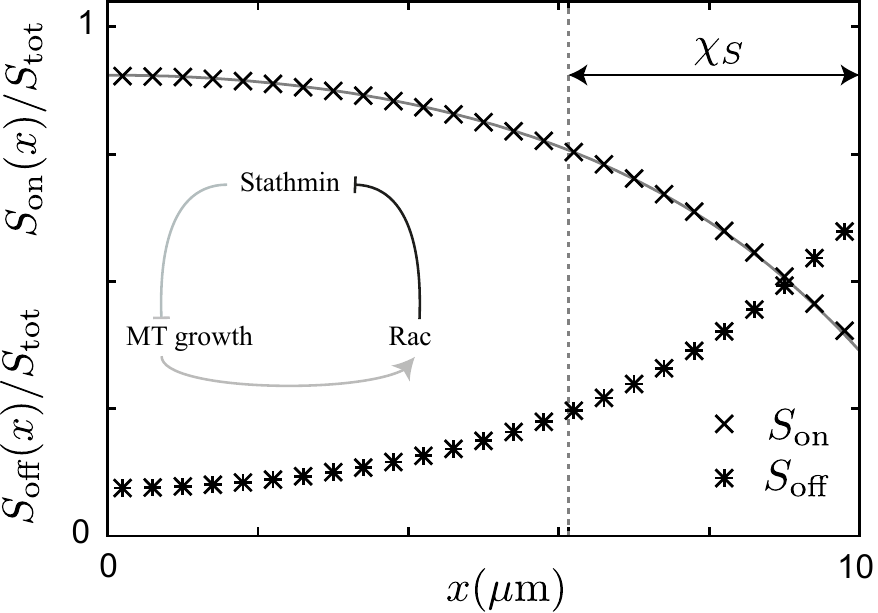}
 \caption{
  Spatial gradient of stathmin activation for constitutively 
 active Rac.  Data points show results from stochastic simulations, 
   lines the analytical result \eqref{eq:son} (for $r_{\text{on}}=1$).
   Parameters are as in Table \ref{tab:parameter} in the main text 
   resulting in 
  $\chi_S\simeq  3.87\,{\rm \mu m}$. 
   There are no fit parameters.
 }
 \label{fig:stathminProfil}
 \end{center}
 \end{figure} 

\section{Gradient in  stathmin activation}
\label{gradient}

The interplay of  stathmin  diffusion and
phosphorylation  by 
active Rac at the cell edge  
establishes a spatial gradient of stathmin
activation, for which we provide a detailed mathematical model 
here.

Stathmin proteins are either in an active  dephosphorylated state
with concentration profile $S_\text{on}(x)$  or in an 
 inactive phosphorylated state with 
concentration profile $S_\text{off}(x)$
(with a total stathmin concentration $S_\text{tot}(x) 
  = S_\text{on}(x)+S_\text{off}(x)$). 
In both states,  stathmin diffuses
freely within the model box $x\in [0;L]$ 
with a constant  diffusion coefficient $D$, which is 
equal for both states. 
  Switching between
 active  and  inactive state
 is a stochastic process with
 rates $k_{\text{on,S}}$ and $k_{\text{off,S}}$.

Moreover, active Rac 
 deactivates stathmin at the cell edge  $x=L$ as  simple second order
reaction with a rate proportional to $r_\text{on}$.
The resulting chemical kinetics for the 
boundary value $S_\text{off}(L)$ 
(strictly speaking the values in the cell-edge region $[L-\delta, L]$) 
including  diffusion
 away from the cell edge can then be described by
\begin{equation}
  \frac{\partial S_\text{off}(L)}{\partial t} = 
  -   \frac{D}{\delta}\left.\frac{\partial S_\text{off}}{\partial x}\right|_{x=L}
   +  k_\text{off,S} r_\text{on} S_\text{on}(L)- k_\text{on,S}S_\text{off}(L).
 \label{eq:appstathL}
\end{equation}

We assume that the phosphatase responsible for 
stathmin activation is homogeneously
distributed within the cytosol
such that stathmin dephosphorylates
everywhere within the box with the constant rate
$k_\text{on,S}$ \citep{Brown1999}
(which depends on  the concentration of the homogeneously
distributed stathmin phosphatase).  
The  distribution of deactivated
stathmin in the one-dimensional box is then 
described by the reaction-diffusion equation 
\begin{equation}
 \frac{\partial S_\text{off}}{\partial t}=
    D\frac{\partial^2 S_\text{off}}{\partial x^2} -
    k_{\text{on,S}}S_\text{off}.
 \label{eq:appdiffstath}
\end{equation}
This equation   is 
  complemented by  the boundary condition \eqref{eq:appstathL}
for $S_\text{off}(L)$
and by a zero diffusive flux condition 
$D\left.\partial S_\text{off}/\partial x\right|_{x=0}=0$ at $x=0$.
The distribution of active stathmin $S_\text{on}(x)$ can be obtained 
analogously (using also $S_\text{tot}(x) = S_\text{on}(x)+S_\text{off}(x)$).

Equations \eqref{eq:appdiffstath} and \eqref{eq:appstathL} 
describe the  stathmin kinetics at the 
mean-field level; equation  \eqref{eq:appstathL} 
 neglects  temporal correlations between 
  Rac and stathmin number fluctuations.
by using the 
mean fraction $r_\text{on}$  in the deactivation process.

In the steady state,
 the time derivatives in  \eqref{eq:appstathL} and 
 \eqref{eq:appdiffstath} vanish,
and the  total stathmin concentration 
$S_\text{tot}= S_\text{on}(x)+S_\text{off}(x) = {\rm const}$  becomes 
 homogeneous due to diffusion.
The corresponding stationary solution of  
\eqref{eq:appstathL} and   \eqref{eq:appdiffstath}
for the  stathmin  activation gradient is 
 \begin{equation}
\frac{S_\text{on}(x)}{S_\text{tot}}=
     1- 2A \cosh(x/\chi_S),  
    \label{eq:appson}
\end{equation}
see Eq.\ \eqref{eq:son} in the main text,
 with the  
 characteristic decay length
\begin{equation}
  \chi_S=\sqrt{{D}/{k_\text{on,S}}}.
\label{eq:appchi}
\end{equation}
The  integration constant  $A$ depends on the 
 degree of stathmin deactivation by Rac at the cell edge $x=L$
 and, thus, on the fraction $r_\text{on}$ 
 of activated Rac. 
 For a fixed  level  $r_\text{on}$  of Rac activation, we find
\begin{equation}
    A= \frac{1}{2} \frac{r_\text{on} k_{\text{off,S}}}{ (D/\delta\chi_S)
         \sinh(L/\chi_S) + 
      (r_\text{on}k_{\text{off,S}}+ k_{\text{on,S}}) \cosh(L/\chi_S) },
 \label{eq:appA}
 \end{equation}
see also Eq.\ \eqref{eq:A} in the main text.

We can compare 
the analytical result \eqref{eq:appson} 
for the  concentration profile of active stathmin at  a 
 fixed  fraction  $r_\text{on}$ of active Rac in Eq.\ \eqref{eq:appA}
with stochastic simulation results for a 
 fixed  level  $r_\text{on}$  of Rac activation
(i.e., keeping $N_R r_\text{on}$ of $N_R$ Rac proteins in the 
simulation permanently in the activated state). 
In  Fig.\ \ref{fig:stathminProfil},
we show this comparison 
for a subsystem with constitutively active Rac corresponding to 
$r_\text{on}=1$ in Eq.\ \eqref{eq:appA} and 
see very good agreement without any adjustable fit parameters.

\begin{figure}[h]
\begin{center}
\includegraphics[width=0.95\linewidth]{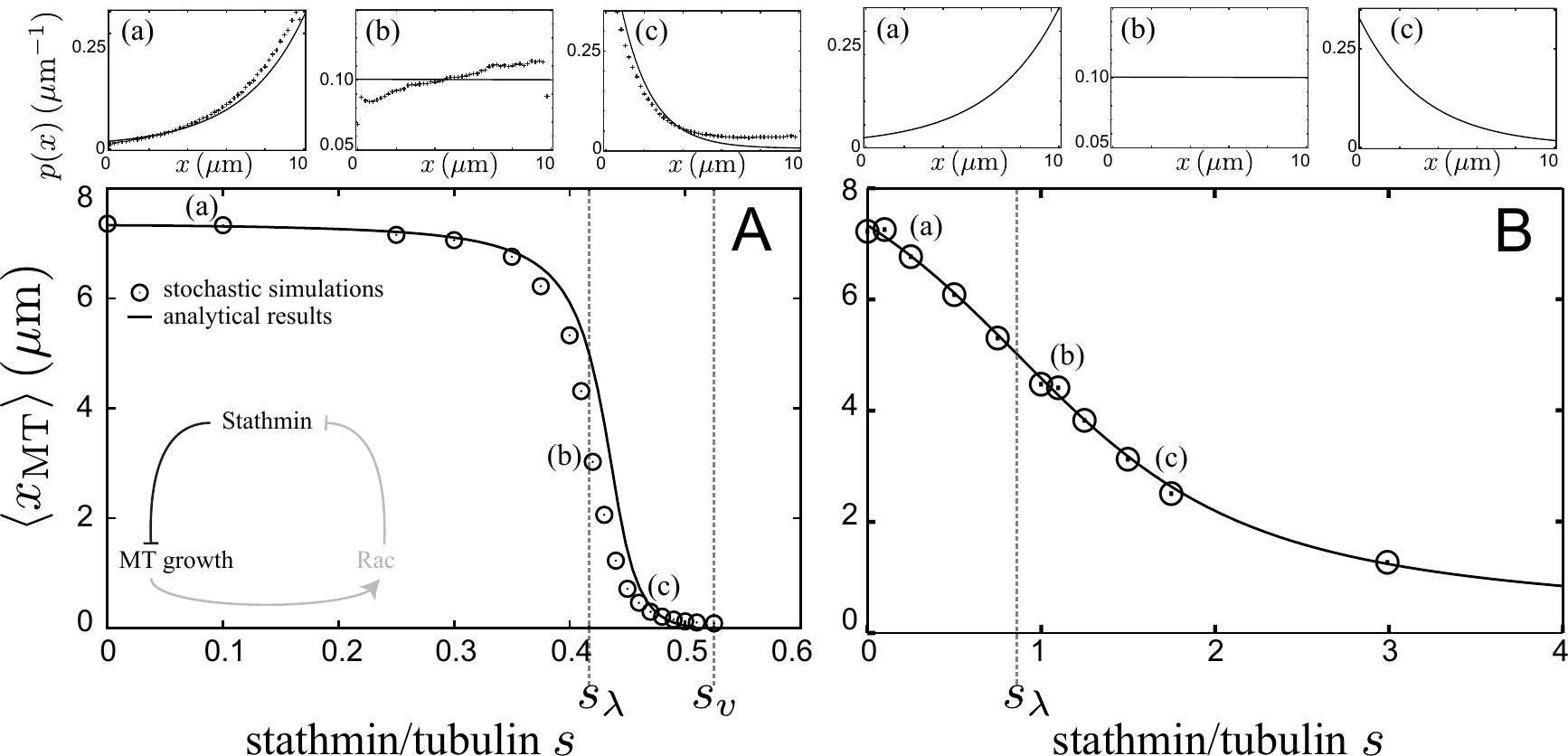}
\caption{ 
  Stochastic simulation data and analytical master equations results for the
  system without Rac  ($r_\text{on}=0$)  and constitutively active 
   stathmin 
 for tubulin-sequestering stathmin (A) and catastrophe-promoting stathmin (B).
We show  results for the
  mean MT length $\langle x_{\text{MT}}\rangle$ in the steady state 
   as a function  of  stathmin/tubulin
   $s=S_{\text{tot}}/{[{T_0}]}$. 
  The black curves  correspond to the analytical result  
  Eq.\ \eqref{eq:xMT} in the main text,
  black symbols to  
   stochastic simulation results.
  The critical concentrations $s_\lambda$ and $s_v$
  are indicated. 
  The insets (a), (b), (c) show MT length distributions for three particular 
  values of $s$, which are also indicated in the main plot, 
  with (a) $s<s_\lambda$, (b) $s=s_\lambda$ and (c) $s>s_\lambda$;
   curves  are  the analytical result  Eq.\ \eqref{eq:pvonx} in the main text,
   symbols are
   stochastic simulation results.
}
\label{fig:Mittelwerte}
\end{center}
\end{figure}

\section{Interrupted feedback for 
 constitutively active stathmin in the 
 absence of Rac}

In this section, 
we present additional results on MT length distributions 
for a subsystem without Rac, i.e., with constitutively active 
stathmin. 

Without Rac, the constitutively active stathmin is homogeneously
distributed, $S_\text{tot} = S_\text{on} = {\rm const}$. 
Therefore, also the length parameter $\lambda$ is 
constant, see  Eq.\ \eqref{lambda} in the main text,
 and the MT mean length 
is given by  Eq.\ \eqref{eq:xMT} in the main text.

Fig.\  \ref{fig:Mittelwerte} shows results for the mean 
MT length $\langle x_{\text{MT}}\rangle$  as a function of 
 stathmin/tubulin
$s = S_\text{tot}/[T_0]$, and insets 
show the corresponding MT length distribution at 
three different stathmin/tubulin $s$.
Fig.\  \ref{fig:Mittelwerte} (A) shows results 
for  tubulin-sequestering stathmin, Fig.\  \ref{fig:Mittelwerte} (B)
for catastrophe-promoting stathmin.
 We find good agreement 
between the analytical result \eqref{eq:xMT} from the main text 
(lines) and stochastic simulation results (data points)
both for tubulin-sequestering and catastrophe-promoting stathmin.

The two critical concentrations $s_\lambda$ and $s_v$ control the mean length 
$\langle x_{\text{MT}}\rangle$, 
as we see in Fig.\  \ref{fig:Mittelwerte}.  
For $s=s_\lambda$, we have $\langle x_{\text{MT}}\rangle=L/2$, for 
$s\ge s_v$, we have $\langle x_{\text{MT}}\rangle=0$ according to 
their definition (see main text).

For tubulin-sequestering stathmin and 
simulation parameters as given in Table \ref{tab:parameter} in the main text 
corresponding to $[T_0] = 19.4 {\rm \mu M}$ or 
a total number of $N_T=10000$ tubulin molecules in the simulation box, 
the critical concentrations of stathmin molecules 
(normalized by the total tubulin concentration $[T_0]$)
are
\begin{equation}
 s_\lambda = \frac{S_\lambda}{[T_0]} = \frac{N_{S,\lambda}}{N_T}=
  0.419
   \quad\text{and}\quad s_v = \frac{S_v}{[T_0]} =\frac{N_{S,v}}{N_T}=
 0.528. 
\label{eq:Ncrit}
\end{equation}
The values $s_\lambda$ and $s_v$ are only weakly $[T_0]$-dependent;
for $[T_0]$ in the range $[T_0] =10-20{\rm \mu M}$, we find 
$s_\lambda = 0.34-0.42$ and $s_v=0.51-0.53$.

For catastrophe-promoting stathmin, 
the MT growth velocity is not affected by stathmin, which  formally results
in an infinite $S_v$. 
A critical concentration $S_\lambda$ can still be defined
in the same way, and  we find 
$S_\lambda = (v_+ \omega_r - v_- \omega_c(0))/k_cv_-$ from Eqs.\ (\ref{lambda})
and (\ref{eq:omegacSon}) in the main text.  
The MT growth velocity 
$v_+$ is linearly increasing with $[T]=[T_0]$, see Eq.\ (\ref{v+})
in the main text, and 
 $\omega_c(0)$ correspondingly decreasing according to Eq.\ (\ref{omegac})
in the main text. 
 For high tubulin concentrations, $s_\lambda$ approaches the 
  $[T_0]$-independent limit
 $s_\lambda \approx \kappa_{\text{on}} d \omega_r/v_-k_c = 0.91$,
 which is significantly above the typical values close to $0.5$ for 
 tubulin-sequestering stathmin.  
For the simulation parameters as given in Table \ref{tab:parameter} 
in the main text corresponding to
$[T_0] = 19.4 {\rm \mu M}$ or 
a total number of $N_T=10000$ tubulin molecules in the simulation box, 
the critical concentration of stathmin molecules 
(normalized by the total tubulin concentration $[T_0]$) is
\begin{equation}
 s_\lambda = \frac{S_\lambda}{[T_0]} = \frac{N_{S,\lambda}}{N_T}=
  0.866.
\label{eq:Ncrit2}
\end{equation}
The value $s_\lambda$ is only weakly $[T_0]$-dependent,
for $[T_0]$ in the range $[T_0] =10-20{\rm \mu M}$, we find 
$s_\lambda = 0.79-0.87$.

The stochastic simulation results for the MT length distribution 
also agree with the analytical result \eqref{eq:pvonx} in the main text.
At the critical stathmin concentration $s=s_\lambda$, 
the condition $\lambda^{-1}=0$  results in a flat 
MT length distribution, 
as confirmed in  inset (b) in Fig.\  \ref{fig:Mittelwerte}. 
For $s>s_{\lambda}$, 
we find the length
distribution of the MT to be a negative exponential
(see inset (c) in Fig.\  \ref{fig:Mittelwerte}).
For  $s<s_{\lambda}$, we find the length
distribution of the MT to be a positive exponential
(see inset (a) in Fig.\  \ref{fig:Mittelwerte}).
For a system which contains constitutively active stathmin and no Rac, 
the growth velocity $v_{+}$ is
 independent of the
position (not taking into account the small  diffusion-induced 
fluctuations of the local stathmin concentration).  
Therefore, we always expect an exponential MT length distribution
according to Eq.\ \eqref{eq:pvonx} in the main text. 
In particular,  we do not expect bimodal MT length 
distributions in the absence of Rac, as confirmed  by 
the stochastic simulation in the insets 
in Fig.\ \ref{fig:Mittelwerte}.

\begin{figure}[h]
 \begin{center}
 \includegraphics[width=0.9\linewidth]{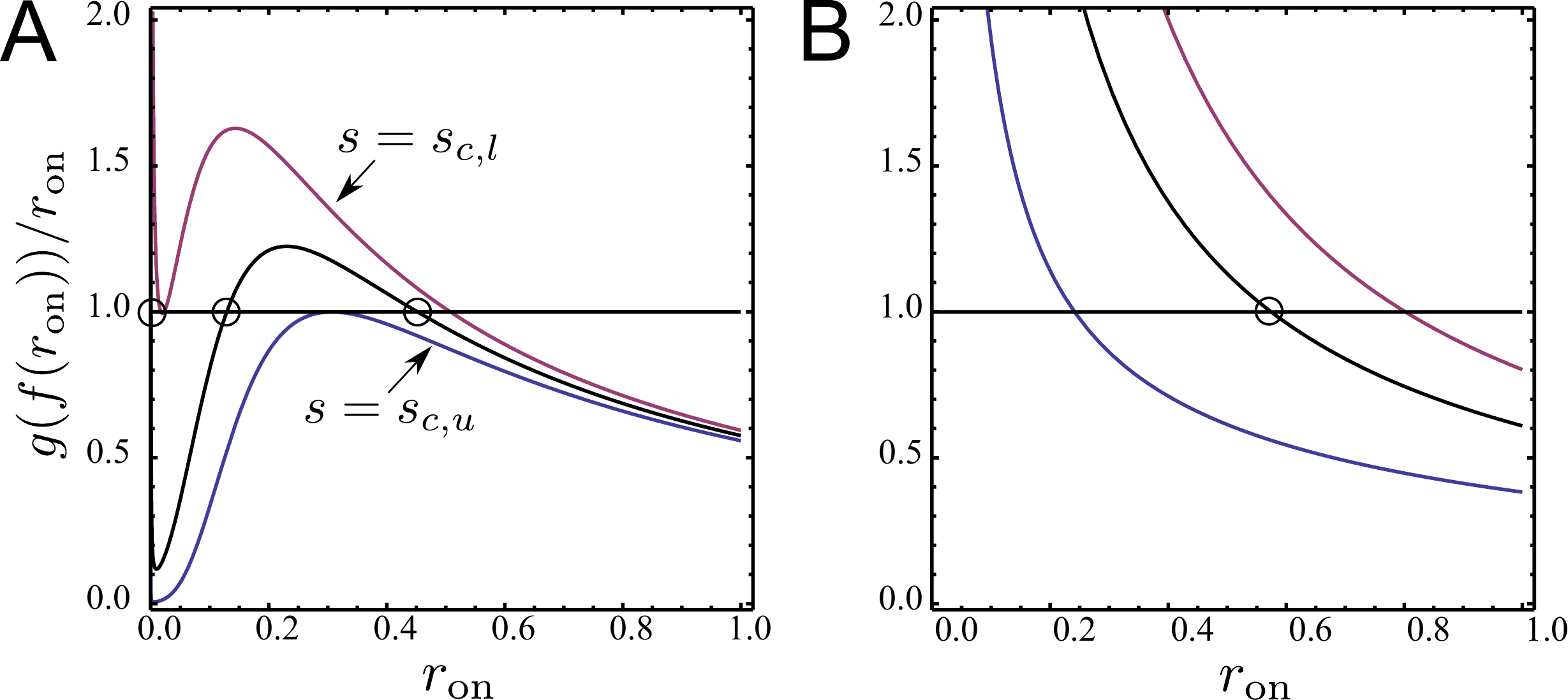}
 \caption{
  Function $g(f(r_\text{on}))/r_\text{on}$ for $m=1$ and three 
   different values of stathmin/tubulin  $s=S_\text{tot}/[T_0]$ 
  for (A) tubulin-sequestering stathmin and (B) catastrophe-promoting stathmin.
  For tubulin-sequestering stathmin (A), the lines are 
  for 
   $s=s_{c,l}=0.446$ (blue),
  $s=s_{c,u}=0.459$ (red), and  $s=0.453$ (black).
   For $s_{c,l} < s < s_{c,u}$ (black line $s=0.453$) 
  there exist three fixed points (circles).
  The middle fixed point is unstable. 
  For catastrophe-promoting stathmin (B) the lines are 
  for    $s=0.1$ (blue),
  $s=1.7$ (red), and  $s=1.1$ (black). 
There is always a single fixed point.
 }
 \label{fig:fixedpoints}
 \end{center}
 \end{figure}

\section{Bifurcation analysis for closed feedback} 

In  the bifurcation analysis 
we consider the stationary system state 
for a closed feedback loop. 

A given fraction $r_\text{on}$ of 
activated Rac then determines the MT contact probability
$p_\text{MT}= m \int_{L-\delta}^L p_+(x,t)$  (see Eq.\ \eqref{eq:pMT}
in the main text)
via (i)  a stationary gradient in stathmin activation (a decreasing 
profile $S_\text{on}(x)$),  (ii) a resulting MT growth velocity 
gradient (an increasing growth velocity $v_x(x)$) established 
   by tubulin sequestering,  and (iii)
 the resulting stationary MT length distribution 
and, in particular, the probability $p_+(x)$ to find a particular  MT 
in the growing state. 
These three dependencies can be described analytically and the 
resulting equations 
can be easily evaluated  numerically:
\begin{itemize}
\item[(i)]
  The stationary gradient in stathmin activation can be calculated 
  by Eq.\ \eqref{eq:son} in the main text, 
  where $r_\text{on}$ enters the gradient 
  amplitude $A$, which is given by Eq.\ \eqref{eq:A} in the main text.
   This allows us to calculate the stathmin 
   activation profile $S_\text{on}(x)/S_\text{tot}$ for any 
  fixed level $r_\text{on}$ of activated Rac. 
 For constitutively active Rac, $r_\text{on}=1$,
   the result is shown and tested versus stochastic simulations 
   in Fig.\ \ref{fig:stathminProfil}.

\item[(ii)]
  From the  stathmin 
   activation gradient $S_\text{on}(x)/S_\text{tot}$
    we obtain $s_\text{on} = S_\text{on}(x)/[T_0] = 
     s (S_\text{on}(x)/S_\text{tot}) $ with the total 
   stathmin/tubulin $s = S_\text{tot}/[T_0]$.
   For tubulin-sequestering stathmin, this is used to determine  the  
 spatially dependent tubulin concentration 
   $[T](x) = [T_0] t(s_\text{on}(x))$  
   through Eq.\ \eqref{tn(x)} in the main text. 
  This free tubulin concentration profile can be used to calculate 
  the MT growth velocity  profile
  $v_+ = v_+([T_0] t(s_\text{on}(x)))$ 
  according to Eq.\ \eqref{eq:v+profile} in the main text.  

\item[(iii)]
  For tubulin-sequestering stathmin, 
  the position-dependent growth velocity $v_+(x)$  also 
  gives rise to a  position-dependent
  catastrophe rate $\omega_c=\omega_c(v_+(x))$
   (see Eq.\ \eqref{omegac} in the main text).
   For catastrophe-promoting stathmin, the stathmin activation
 gradient  directly gives rise to  position-dependent
  catastrophe rate $\omega_c=\omega_c(S_\text{on}(x))$ 
   via relation \eqref{eq:omegacSon} in the main text. 
  Finally, we obtain a position dependent MT growth parameter 
  $\lambda(x) = v_+(x)v_-/(v_+(x)\omega_r-v_-\omega_c(x))$ for both 
  models of stathmin action
   (see Eq.\ \eqref{lambda} in the main text). 
  This allows us to calculate the relevant MT length distribution 
   $p_+(x)$ according to Eq.\ \eqref{eq:pgrad} from the main text.
 
\end{itemize}

This scheme allows us to calculate $p_+(x)$ and, thus, $p_{MT}$ for 
 any fixed level $r_\text{on}$, i.e., to implement a function 
  $p_\text{MT} = f(r_\text{on})$. Vice versa, for a closed feedback, 
   the contact probability also determines the  Rac activation level 
 $r_\text{on}$ via \eqref{eq:ronstat} in the stationary state,
 which specifies a second function $r_\text{on} = g(p_\text{MT})$. 
 At the stationary state for closed feedback, the fixed point 
  condition  
   $r_\text{on} = g(f(r_\text{on}))$ has to hold, which selects 
   possible fixed point values $r_\text{on}^*$
  for the Rac activation level $r_\text{on}$. 
  If 
\begin{equation}
  \left. \frac{dg(f(r))}{dr}\right|_{r=r_\text{on}^*} <1
\end{equation}
the fixed point is stable, otherwise it is unstable. 
For a stable fixed point, an increase  $\delta r_\text{on}$
in $r_\text{on}$ by  perturbation
gives rise to a down-regulation of $r_\text{on}$ because the 
corresponding increase $\delta p_\text{MT}$ is 
not sufficient to maintain the increased level 
$r_\text{on}+\delta r_\text{on}$.

Analyzing the function $g(f(r_\text{on}))$ 
for tubulin-sequestering stathmin, we find 
two saddle-node bifurcations typical for a bistable switch.
For stathmin/tubulin $s=S_\text{tot}/[T_0]$ between a
lower critical value $s_{c,l}$ ($s_{c,l}=0.446$ for $m=1$ increasing to 
 $s_{c,l} = 0.453$ for $m=10$) and an upper critical value 
$s_{c,u}$ ($s_{c,u}=0.459$ for $m=1$ increasing to 
 $s_{c,u} = 0.492$ for $m=10$ and approaching $s_{c,u} \approx
s_v = 0.528$ for large $m$) there exist three fixed points, 
 the middle one of which is unstable  (see 
Fig.\ \ref{fig:fixedpoints} A).

\begin{figure}[h]
 \begin{center}
 \includegraphics[width=0.99\linewidth]{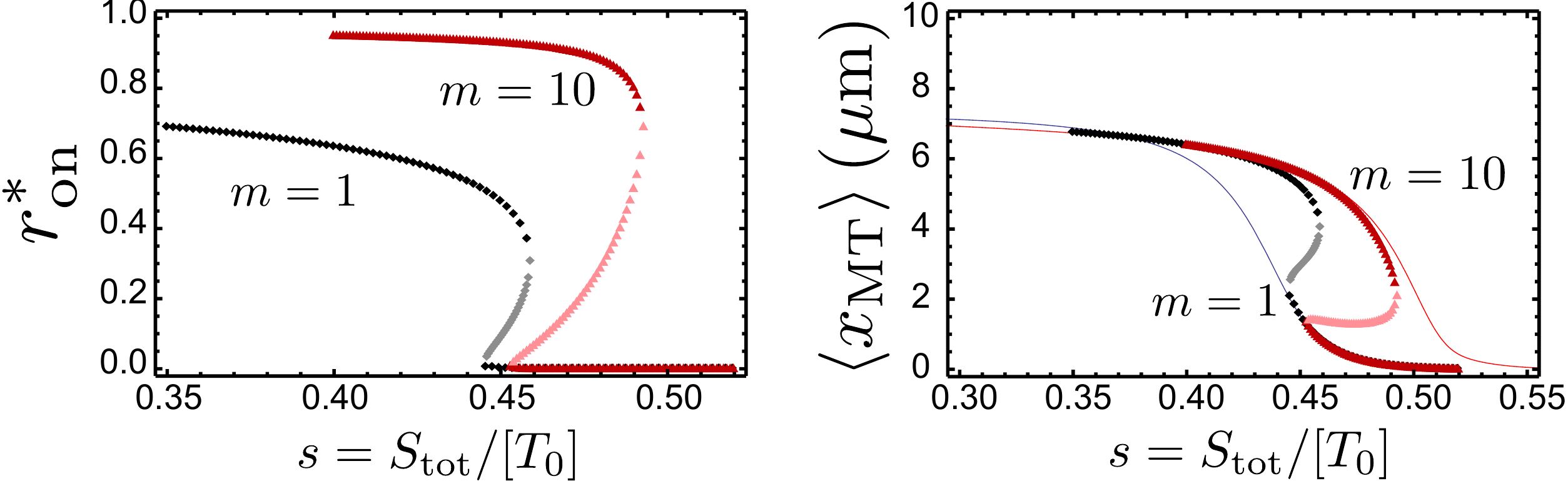}
 \caption{
 Left: Rac activation fixed points $r_\text{on}^*$ 
  for tubulin-sequestering stathmin as a function of 
   stathmin/tubulin 
     $s=S_\text{tot}/[T_0]$  for $m=1$ (black data points) and $m=10$ 
  (red data points). 
   For $s_{c,l} < s < s_{c,u}$ there exist three fixed points.
  The middle fixed point are unstable (light red and light black data
  points). 
  Right: Corresponding average MT length $\langle x_\text{MT} \rangle$ 
   for the fixed point values  $r_\text{on}^*$ as a function of $s$. 
  Lines show the  average MT length $\langle x_\text{MT} \rangle$ 
  for a system without Rac and 
   constitutively active stathmin   ($r_\text{on}=0$, blue line)
   and for constitutively active Rac ($r_\text{on}=1$, red line). 
 }
 \label{fig:bifurcation}
 \end{center}
 \end{figure}

This bifurcation behavior represents a bistable switch with 
stathmin/tubulin $s$ as control parameter as can be 
clearly seen from Fig.\ \ref{fig:bifurcation}.
In the left Fig.\ \ref{fig:bifurcation}, 
  the Rac activation fixed points $r_\text{on}^*$ are shown as a function
of $s$. In the right Fig.\ \ref{fig:bifurcation}, the corresponding average  
  MT length $\langle x_\text{MT} \rangle$ is shown.

For catastrophe-promoting stathmin we always find a 
{\em single} fixed point and no sign of a bifurcation
for all stathmin/tubulin  $s=S_\text{tot}/[T_0]$ values
(see Fig.\ \ref{fig:fixedpoints} B).

\section{Robustness of results} 

In this section, we address the robustness of our results with respect 
to changes in the catastrophe model and the system length.

\begin{figure}[h!]
\begin{center}
\includegraphics[width=0.95\linewidth]{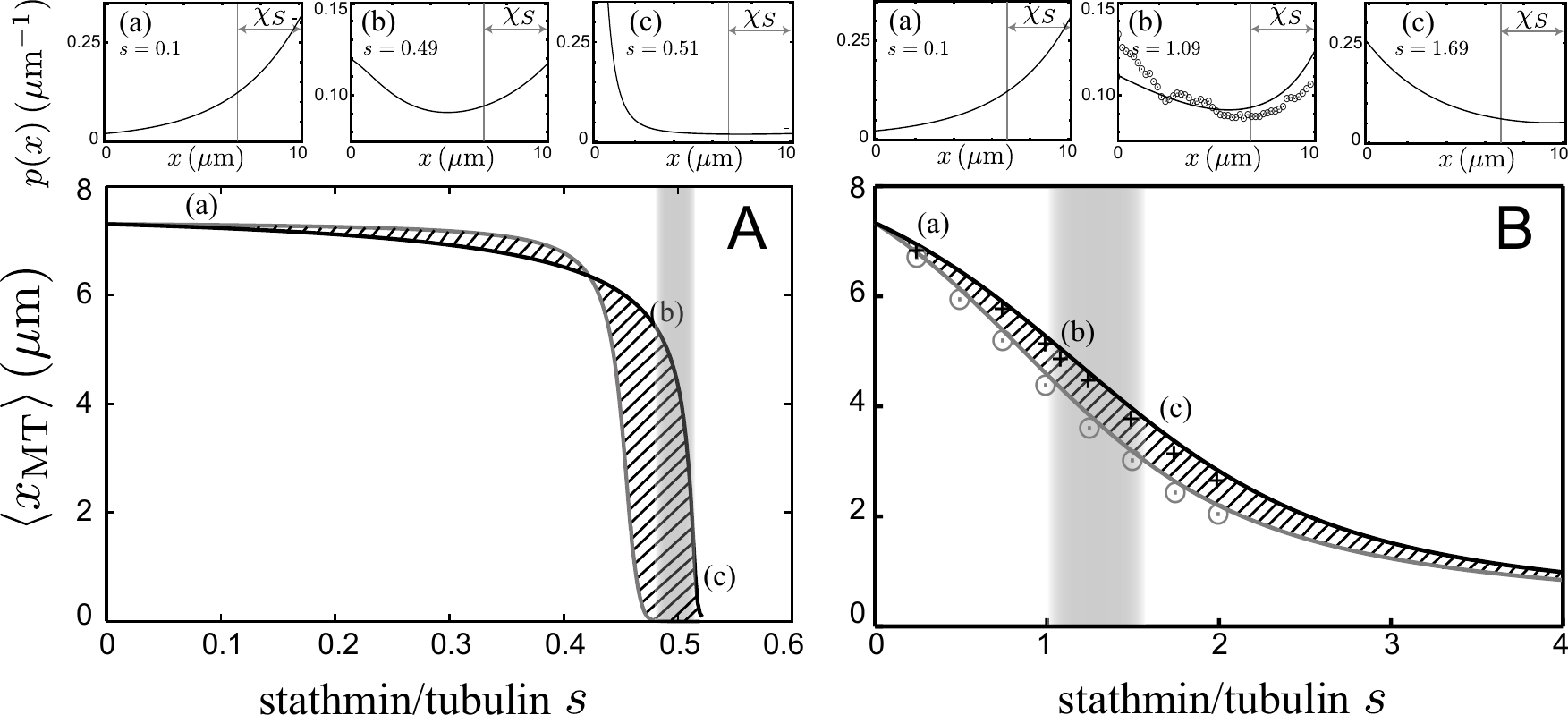}
\caption{ 
  Results for the alternative Flyvbjerg catastrophe model
  analogously to Fig.\ \ref{fig:MittelwerteallesRac} in the main text. 
 Stochastic simulation data (data points) 
  and analytical master equation  results  (solid lines)
   for the
  mean MT length $\langle x_{\text{MT}}\rangle$ 
   as a function  of stathmin/tubulin
    $s=S_{\text{tot}}/{[{T_0}]}$.
  We compare the system with 
  constitutively active Rac ($r_\text{on}=1$, black lines and crosses) 
  in comparison to the  system without Rac ($r_\text{on}=0$, gray lines and 
   circles) both 
 for tubulin-sequestering stathmin (A) and catastrophe-promoting stathmin
  (B).
    The hatched area indicates the possible MT length gain by Rac regulation.
  The gray shaded area indicates the
  region, in which MTs exhibit bimodal length distributions
  for constitutively active Rac.
  The insets (a), (b), (c) show the corresponding 
  MT length distributions for three particular 
  values of $s$ with (a) $s<s_\lambda$, (b) $s=s_\lambda$ and (c)
  $s>s_\lambda$.
}
\label{fig:Flyvbjerg}
\end{center}
\end{figure}

\begin{figure}[h]
 \begin{center}
 \includegraphics[width=0.99\linewidth]{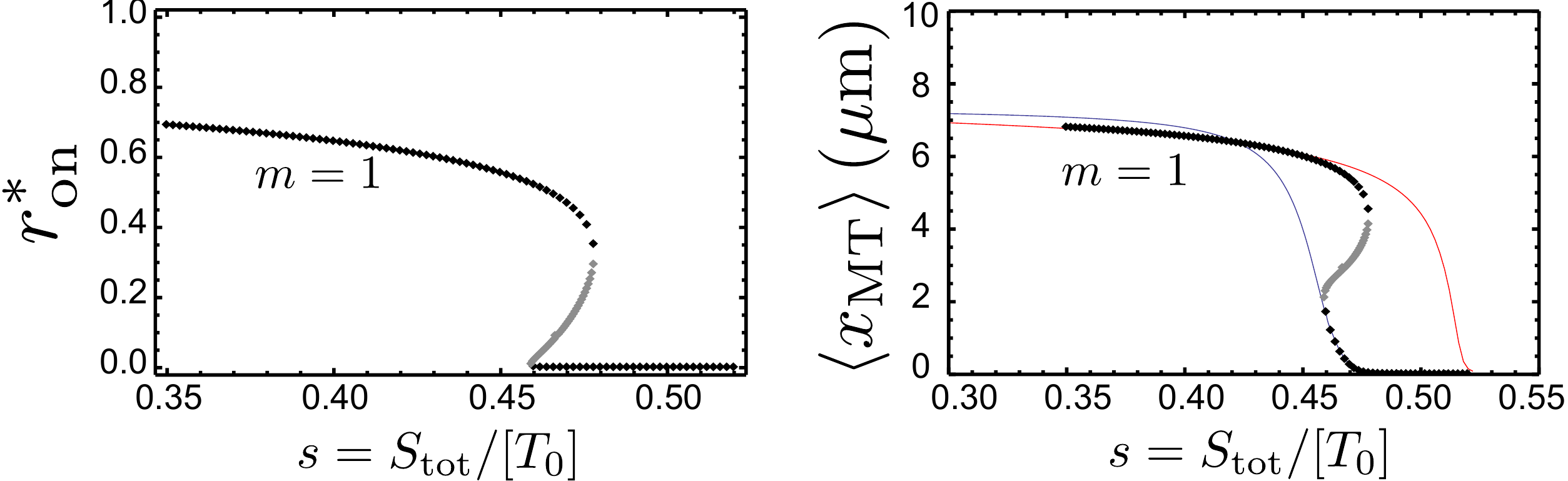}
 \caption{
  As in Fig.\ \ref{fig:bifurcation} but for the 
  Flyvbjerg catastrophe mode. 
 Left: Rac activation fixed points $r_\text{on}^*$ 
  for tubulin-sequestering stathmin as a function of 
   stathmin/tubulin  $s=S_\text{tot}/[T_0]$  for $m=1$. 
   For $s_{c,l} < s < s_{c,u}$ there exist three fixed points.
  The middle fixed point are unstable (light black data
  points). 
  Right: Corresponding average MT length $\langle x_\text{MT} \rangle$ 
   for the fixed point values  $r_\text{on}^*$ as a function of $s$. 
  Lines show the  average MT length $\langle x_\text{MT} \rangle$ 
  for a system without Rac and 
   constitutively active stathmin   ($r_\text{on}=0$, blue line)
   and for constitutively active Rac ($r_\text{on}=1$, red line). 
 }
 \label{fig:bifurcation_Flyv}
 \end{center}
 \end{figure} 

\subsection{Robustness with respect to the catastrophe model}

The catastrophe model described by Eq.\ 
(\ref{omegac})  in the main text and used throughout the manuscript
is based on the experimental  results by Janson {\it et al.}  
\citep{Janson2003b} and, in this sense, of phenomenological nature. 
Other catastrophe models have been formulated in the literature. 
Because there is no strict consensus on a particular catastrophe 
model, it is important that results are robust with respect to 
 a change of the catastrophe model.

Another  frequently used catastrophe model due to Flyvbjerg {\it et al.} 
is based on an analytical  calculation of the first passage rate 
 to a state with vanishing GTP-cap  for  a 
model for cooperative hydrolysis of GTP-tubulin \citep{Flyvbjerg1996}.
In a cooperative model, hydrolysis proceeds  by 
 a combination of both  random  and  vectorial
mechanisms \citep{Li2009}. 
 In Ref.\ \citep{Flyvbjerg1996}, the catastrophe 
rate has been calculated 
as implicit function of the growth velocity  
$v_{+}$ and two  hydrolysis parameters 
 $v_h$ (characterizing the vectorial part) and $r$
(characterizing the random part).
 The exact dimensionless catastrophe rate 
$\alpha \equiv \omega_{c} D^{-1/3}r^{-2/3}$ is 
given by the smallest solution of 
\begin{equation}
   Ai'(\gamma^2-\alpha) = -\gamma Ai(\gamma^2-\alpha)
\label{Ai}
\end{equation}
with $\gamma \equiv vD^{-2/3}r^{-1/3}/2$,
where $v\equiv v_+-v_h$ and $D\equiv (v_{+}+v_h)d/2$ 
[$Ai'(x) \equiv dAi(x)/dx$].
Here $Ai$ denotes the first Airy function. 
We use a numerical implementation of this
analytical result
in simulations and mean-field calculations:
We solve Eq.\ (\ref{Ai}) numerically to calculate the function 
$\alpha = \omega_{c} D^{-1/3}r^{-2/3}$ as a function of $\gamma$.
From this numerical solution  we obtain the catastrophe 
rate as function of the MT growth velocity,  
$\omega_c = \omega_c(v_+)$.
The hydrolysis parameters $v_h\simeq 4.2\times 10^{-9}\;{\rm m/s}$ and 
$r\simeq 3.7\times 10^6\, {\rm m}^{-1}{\rm s}^{-1}$ \citep{Flyvbjerg1996} are
fixed during the simulation.

Results for this alternative catastrophe model 
by Flyvbjerg  {\it et al.}  are shown in Fig.\ \ref{fig:Flyvbjerg}.
Analogously to Fig.\ \ref{fig:MittelwerteallesRac} in the main text, 
we show both results for tubulin-sequestering stathmin 
(Fig.\ \ref{fig:Flyvbjerg} A)
and catastrophe-promoting stathmin (Fig.\ \ref{fig:Flyvbjerg} B).
We obtain the same main features using the alternative catastrophe model:
We find  a switchlike dependence of MT length on the  overall
stathmin/tubulin both for tubulin-sequestering and 
catastrophe-promoting stathmin.
We obtain 
bimodal MT length distributions both for tubulin-sequestering 
and catastrophe-promoting stathmin.
We find 
bistability for tubulin-sequestering stathmin (see Fig.\ 
\ref{fig:bifurcation_Flyv}), whereas
there is no bistability for  catastrophe-promoting stathmin.

\begin{figure}[h!]
\begin{center}
\includegraphics[width=0.95\linewidth]{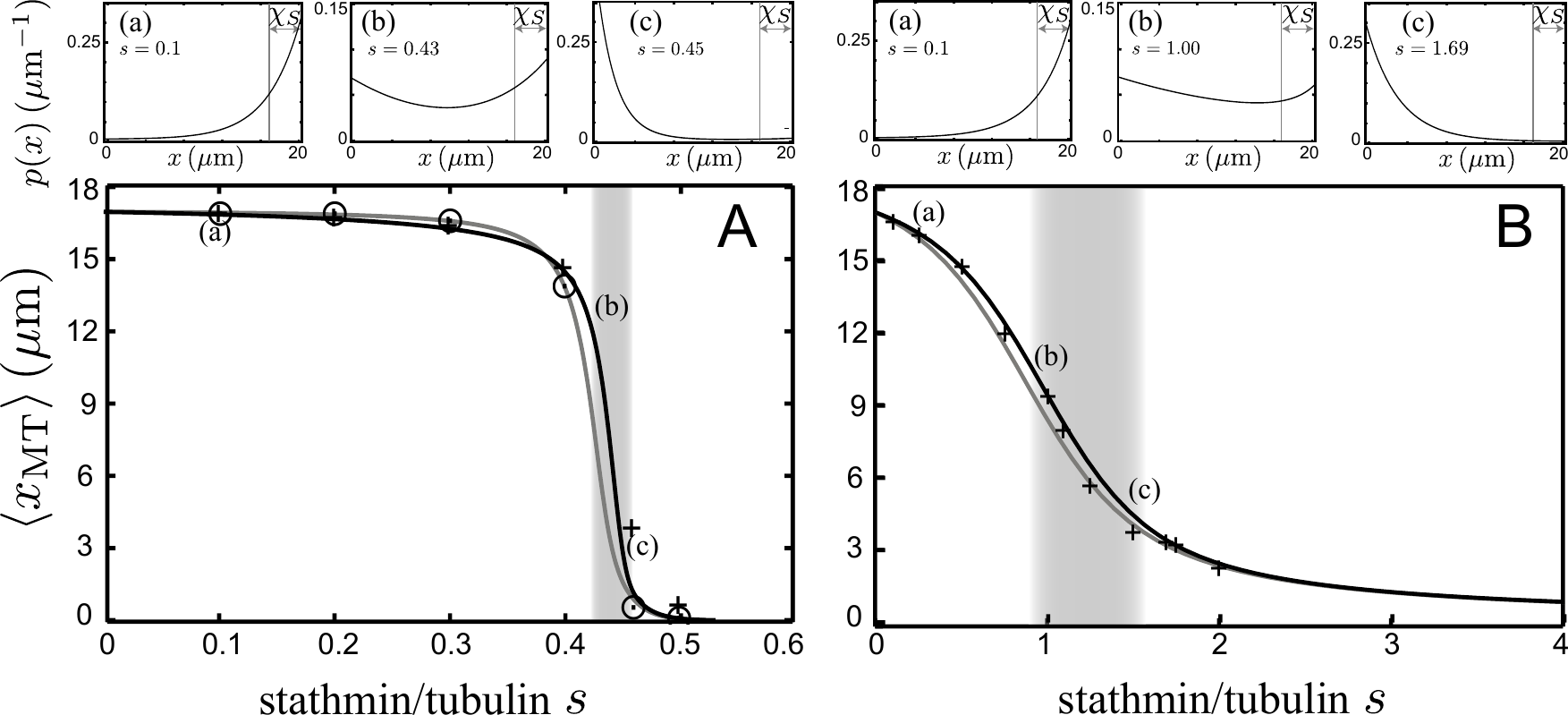}
\caption{ 
  Results for a larger system of length $L=20\,{\rm \mu m}$, 
  analogously to Fig.\ \ref{fig:MittelwerteallesRac} in the main text,
  which is for  $L=10\,{\rm \mu m}$. 
 Stochastic simulation data (data points) 
  and analytical master equation  results  (solid lines)
   for the
  mean MT length $\langle x_{\text{MT}}\rangle$ 
   as a function  of stathmin/tubulin
    $s=S_{\text{tot}}/{[{T_0}]}$.
  We compare the system with 
  constitutively active Rac ($r_\text{on}=1$, black lines and crosses) 
  in comparison to the  system without Rac ($r_\text{on}=0$, gray lines and 
   circles) both 
 for tubulin-sequestering stathmin (A) and catastrophe-promoting stathmin
  (B).
  The gray shaded area indicates the
  region, in which MTs exhibit bimodal length distributions
  for constitutively active Rac.
  The insets (a), (b), (c) show the corresponding 
  MT length distributions for three particular 
  values of $s$ with (a) $s<s_\lambda$, (b) $s=s_\lambda$ and (c)
  $s>s_\lambda$.
}
\label{fig:20}
\end{center}
\end{figure}

\subsection{Robustness with respect to the system length}

Cells have different lengths. Therefore we also investigated 
the robustness of our results with respect to changes in the system
length $L$. In the main text we used $L=10\,{\rm \mu m}$, 
here we present additional simulation results 
for a longer system $L=20\,{\rm \mu m}$ in Fig.\ \ref{fig:20}. 
It is important to notice that one {\em fixed} length scale 
within the feedback loop is set by the characteristic scale 
$\chi_S=\sqrt{{D}/{k_\text{on,S}}}$ (see Eq.\ (\ref{eq:appchi})) 
of the stathmin activation gradient. 
For parameters as in Table \ref{tab:parameter} in the main text, we have
  $\chi_S\simeq  3.87\,{\rm \mu m}$. 
The length scale $\chi_S$ 
 is the typical size of the cell-edge region  $L-\chi_S<x<L$
in which stathmin is deactivated. 
The stathmin activation profile becomes very flat
in the remaining region  $0< x < L-\chi_S$, which increases 
in size if  the system size $L$ is increased.

Qualitatively, we find 
the same behavior for a longer system $L=20\,{\rm \mu m}$  
as for the shorter 
system $L=10\,{\rm \mu m}$ 
 with  a switchlike dependence of MT length on the  overall
stathmin/tubulin, a  bimodal MT length distributions 
both for tubulin-sequestering and 
catastrophe-promoting stathmin and 
bistability only  for tubulin-sequestering stathmin.
As a result of the more shallow gradient in large parts of the cell,
however, 
the switching behavior of the MT length as a function of the 
total stathmin level becomes steeper, see Fig.\ \ref{fig:20}
for tubulin-sequestering stathmin.
Accordingly, for tubulin-sequestering stathmin, 
the windows of stathmin concentrations, 
where we find bimodal MT length distributions 
and where we find bistability, 
are narrower for longer cells.


\end{document}